\documentclass[review]{elsarticle}
\usepackage{}

\usepackage{lineno,hyperref}
\usepackage{amssymb}
\usepackage{color}
\usepackage{amsmath}
\usepackage{tikz}
\usepackage{amsfonts}
\usepackage{mathrsfs}
\usepackage{amsthm}

\usepackage{bbding}
\usepackage{mathrsfs}
\usepackage{amsmath, amsfonts, amssymb, mathrsfs, txfonts}
\usepackage{graphicx, subfigure}
\usepackage{wasysym}
\usepackage{color}
\usepackage{bm}
\usepackage{enumerate}

\usepackage{pifont}
\usepackage{txfonts}
\usepackage{amsmath}
\usepackage{graphicx}
\usepackage{amsfonts}
\usepackage{amssymb}
\usepackage{mathrsfs,psfrag,eepic,epsfig}
\usepackage{epstopdf}

\newtheorem{thm}{Theorem}[section]

 \newtheorem{prop}[thm]{Proposition}
 \theoremstyle{definition}
 
 \theoremstyle{remark}

\biboptions{numbers,sort&compress}

\modulolinenumbers[5]

\journal{Journal of \LaTeX\ Templates}

\makeatletter \@addtoreset{equation}{section}
\renewcommand{\theequation}{\arabic{section}.\arabic{equation}}

\begin{document}

\begin{frontmatter}
\title{Riemann-Hilbert approach to the inhomogeneous fifth-order nonlinear Schr\"{o}dinger equation with non-vanishing boundary conditions}
\tnotetext[mytitlenote]{Project supported by the Fundamental Research Fund for the Central Universities under the grant No. 2019ZDPY07.\\
\hspace*{3ex}$^{*}$Corresponding author.\\
\hspace*{3ex}\emph{E-mail addresses}: jinjieyang@cumt.edu.cn (J.J. Yang), sftian@cumt.edu.cn and
shoufu2006@126.com (S. F. Tian),
zqli@cumt.edu.cn (Z. Q. Li)}

\author{Jin-Jie Yang, Shou-Fu Tian$^{*}$ and Zhi-Qiang Li}
\address{
School of Mathematics and Institute of Mathematical Physics, China University of Mining and Technology,\\ Xuzhou 221116, People's Republic of China\\
}

\begin{abstract}
We consider the inhomogeneous fifth-order nonlinear Schr\"{o}dinger (ifoNLS) equation with nonzero boundary condition in detailed. Firstly, the spectral analysis of the scattering problem is carried out. A Riemann surface and affine parameters are first introduced to transform the original spectral parameter to  a new parameter in order to avoid the multi-valued problem. Based on Lax pair of the ifoNLS equation, the Jost functions are obtained, and their analytical, asymptotic, symmetric properties, as well as the corresponding properties of the scattering matrix are established systematically. For the inverse scattering problem, we discuss the cases that the scattering coefficients have simple zeros and double zeros, respectively, and we further derive their corresponding exact solutions. Moreover, some interesting phenomena are found when we choose some appropriate parameters for these exact solutions, which is helpful to study the propagation behavior of these solutions.
\end{abstract}

\begin{keyword}
The inhomogeneous fifth-order nonlinear Schr\"{o}dinger equation \sep  Nonzero boundary condition \sep Simple zeros and double zeros \sep Riemann-Hilbert problem \sep Soliton solutions.
\end{keyword}

\end{frontmatter}


\section{Introduction}
In the past few decades, lots of theoretical and experimental studies have been carried out on the model of distributed nonlinear Schr\"{o}dinger (NLS) equation with time-dependent complicated potential, spatially/temporally cubic or
cubic-quintic nonlinearities \cite{BBJ-2007, PG-1998}. The standard NLS equation, i.e.,
\begin{align}\label{nls}
i\psi_{t}+\frac{1}{2}\psi_{xx}+|\psi|^{2}\psi=0,
\end{align}
is a completely integrable appearing in various physical systems including plasma physics, nonlinear optics, Bose-Einstein condensation and the hydrodynamics, etc. In addition, it possesses many kinds of soliton solutions: rouge wave, breather wave, bright and  dark soliton solutions. It is worth noting that the constant coefficient equation is highly idealized in the physical background. Therefore, the inhomogeneity of media should be considered when describing physical phenomena. From deforming the inhomogeneous Heisenberg ferromagnetic system, Radha \cite{RR-2007} proposed  the inhomogeneous fifth-order NLS equation
\begin{align}\label{RR1}
\begin{split}
iu_{t}-i\epsilon u_{xxxxx}-10i&\epsilon|u|^{2}u_{xxx}-20i\epsilon u_{x}u^{*}u_{xx} -
10i\epsilon(|u_{x}|^{2}u)_{x}-30i\epsilon|u|^{4}u_{x}\\&+(fu)_{xx}+2u\left(f|u|^{2}
+\int_{-\infty}^{x}f_{x}|u|^{2}dx'\right)-i(hu)_{x}=0,
\end{split}
\end{align}
where $u=u(x,t)$ is a complex variable, the linear functions $f$ and $h$ about the variable $x$  denote the inhomogeneities in the medium, that is $f=m_{1}x+n_{1}$ and $h=m_{2}x+n_{2}$ with the real constant $m_{1}, m_{2}, n_{1}, n_{2}$ and $\epsilon$, and the star represents the complex conjugate.
In this work, we mainly investigate the following simplified inhomogeneous fifth-order nonlinear Schr\"{o}dinger (ifoNLS) equation for $f=h\equiv 1$
\begin{align}\label{Q1}
\begin{split}
iu_{t}-i\epsilon u_{xxxxx}-10i\epsilon|u|^{2}u_{xxx}-&20i\epsilon u_{x}u^{*}u_{xx}-30i\epsilon|u|^{4}u_{x}\\&-
10i\epsilon(|u_{x}|^{2}u)_{x}+u_{xx}+2u|u|^{2}-iu_{x}=0.
\end{split}
\end{align}
The ifoNLS equation can be regarded as an extension of NLS equation \eqref{nls}, which has been studied in many respects, such as solitary wave, breather wave and rogue wave solutions via Darboux transformation \cite{FengLL-2019}. Yinman \cite{Yinnan-2015} has obtained the rouge wave solutions, and further discussed the evolution influence of the rouge wave solutions with different parameters by using these solutions.

It follows that via the transformation $u=qe^{2iq_{0}^{2}t}$, the Eq. \eqref{Q1} can be written as
\begin{align}\label{Q2}
\begin{split}
iq_{t}+2(|q|^{2}-q_{0}^{2})q-i\epsilon q_{xxxxx}-&10i\epsilon|q|^{2}q_{xxx}-20i\epsilon q_{x}q^{*}q_{xx}\\&-30i\epsilon|q|^{4}q_{x}-
10i\epsilon(|q_{x}|^{2}q)_{x}+q_{xx}-iq_{x}=0.
\end{split}
\end{align}
The purpose in the present work is devoted to study solutions to Eq. \eqref{Q2} with the nonzero boundary condition at infinity, that is,
\begin{align}\label{Q3}
\lim_{x\rightarrow\pm\infty}q(x,0)=q_{\pm},
\end{align}
here $q_{\pm}$ are complex constants and $|q_{\pm}|=q_{0}\neq0$. It is worth noting that the  advantage of Eq. \eqref{Q2} over  Eq. \eqref{Q1} is that Eq. \eqref{Q2} satisfying the boundary condition Eq. \eqref{Q1} is time independent when $x$ tends to infinity. As we known, the soliton solutions to ifoNLS equation  have not been reported by the Riemann-Hilbert (RH) probelm  \cite{Shabat-1976,Zakharov-1984} with the nonzero boundary condition. The RH method is a powerful tool for solving integrable partial differential equations. After further development of  this method, it is not only limited to solve soliton solutions \cite{Ablowitz-1991}-\cite{RHP-11} and long-time asymptotic behavior \cite{longtime-0}-\cite{longtime-7}  under zero boundary conditions (ZBC), but also to further solve soliton solutions \cite{NZBC-1}-\cite{NZBC-16} and long-time asymptotic behavior \cite{NZBCtime-1}-\cite{NZBCtime-5} under non-zero boundary conditions (NZBC), time-periodic boundary condition \cite{TZ-2018}, shock problem \cite{shock-2007}, step-like
initial data \cite{step-2011}, etc. 

The main purpose of the work is to construct the soliton solutions of the ifoNLS equation under the condition of NZBC Eq. \eqref{Q3} by using RH method.  It is noted that the propagation behavior of the final solution of the equation is different from that of the case in \cite{NZBC-14}, including the inability to obtain the Kuznetsov-Ma soliton solution and the  Akhmediev breather solution. The solution obtained in the work is not parallel to the $x$-axis or the $t$-axis which is the so-called non-stationary solitons. 

The main results of the work are the following theorems:
\begin{thm}
The solution of the Eq. \eqref{Q1} with simple zeros under the reflection-less potential can be written as
\begin{align}\label{Jie-Q}
q(x,t)=-q_{-}+i\frac{\det\left(\begin{array}{cc}
                      \mathcal {G} & \varrho \\
                      \varpi^{T} & 0
                    \end{array}\right)}{\det \mathcal {G}},
\end{align}
where the elements of the matrix are defined by Eq. \eqref{xingshi} and Eq. \eqref{Q31}, the discrete spectrum set $\mathbb{Z}=\{\eta_{n},\hat{\eta}_{n}\}$, and the elements $b_{-}(z_{n})$ and $b_{+}(z_{n}^{*})$ are constant.
\end{thm}
\begin{thm}
The solution of the Eq. \eqref{Q1} with double zeros under the reflection-less potential can be given by
\begin{align}\label{T20}
q(x,t)=-q_{-}+i\frac{\det\left(
                \begin{array}{cc}
                  \tilde{G} & \nu \\
                  \omega^{T} & 0 \\
                \end{array}
              \right)}{\det \tilde{G}},
\end{align}
where the elements of the matrix are determined by Eqs. \eqref{4.14}, \eqref{4.19} and \eqref{T17}, the discrete spectrum set $\mathbb{Z}=\{\eta_{n},\hat{\eta}_{n}\}$, and the elements $b_{-}(z_{n})$ , $b_{+}(z_{n}^{*})$,  $d_{-}(z_{n})$ and $d_{+}(z_{n}^{*})$ are constant.
\end{thm}
The structure of this work is as follows: In section 2, the direct scattering process is presented, including spectrum analysis, plane transformation, analyticity and symmetry of Jost functions and scattering matrix, discrete spectrum, and the establishment of residue conditions. In section 3, we successfully construct the RH problem of the ifoNLS equation with simple zeros, and derive the exact formulae of the solutions to the Eq. \eqref{Q1} with reflection-less potential. In section 4, the ifoNLS equation with double zeros is discussed,  we then study the residue condition generated by double zeros. Similar to the case of simple zeros with many different processes,  the exact solution expression of Eq. \eqref{Q1}  including trace formula and theta condition, is  presented in the appendix B. Finally, some conclusions and some conjectures of higher order zeros are given in the last section.

\section{The direct scattering}
\subsection{The Lax pairs}
From \cite{RR-2007}, the Lax pairs of the ifoNLS Eq. \eqref{Q2} can be read as
\begin{align}\label{Q4}
&\left\{\begin{aligned}
\phi_{x}&=U\phi=(-ik\sigma_{3}+U_{0})\phi,\\
\phi_{t}&=V\phi=\left(\begin{array}{cc}
V_{11} & V_{12}\\
-V^{*}_{12} & -V_{11}
\end{array}\right)\phi,
     \end{aligned}  \right.
\end{align}
where $U_{0}=\left(\begin{array}{cc}
    0  & q \\
    -q^{*} & 0  \\
  \end{array}\right),\qquad
  \sigma_{3}=\left(\begin{array}{cc}
    1  & 0 \\
    0 & -1  \\
  \end{array}\right)$\quad and
\begin{align*}
&V_{11}=-16ik^{5}\epsilon+8ik^{3}|q|^{2}\epsilon+4k^{2}\epsilon(qq_{x}^{*}-
q_{x}q^{*})-ik+i|q|^{2}-2ik^{2}-6ik\epsilon|q|^{4}
\\&-2ik\epsilon(qq_{xx}^{*}+q_{xx}q^{*}-|q_{x}|^{2})
+\epsilon(q_{xxx}q^{*}-qq_{xxx}^{*}+q_{x}q_{xx}^{*}-q_{xx}q_{x}^{*}
+6|q|^{2}q^{*}q_{x}-6|q|^{2}q_{x}^{*}q),\\
&V_{12}=16k^{4}\epsilon q+8ik^{3}q_{x}\epsilon-4k^{2}\epsilon(q_{xx}+2|q|^{2}q)-
2ik\epsilon(q_{xxx}+6|q|^{2}q_{x})+2kq+iq_{x}+q\\
&+\epsilon(q_{xxxx}+8|q|^{2}q_{xx}
+2q^{2}q_{xx}^{*}+4|q_{x}|^{2}q+6|q_{x}|^{2}q^{*}+6|q|^{4}q),
\end{align*}
and $k\in\mathbb{C}$ is a spectral parameter. By using the NZBC Eq. \eqref{Q3}, the Lax pairs can be written as
\begin{align}\label{Q5}
\phi_{x}=X_{\pm}\phi,\qquad \phi_{t}=T_{\pm}\phi,
\end{align}
here
\begin{align}\label{Q6}
 &\left\{\begin{aligned}
X_{\pm}&=\lim_{x\rightarrow\pm\infty} U=-ik\sigma_{3}+Q_{\pm},\\
T_{\pm}&=\lim_{x\rightarrow\pm\infty} V=\left(16k^{4}\epsilon-8k^{2}q_{0}^{2}\epsilon
+2k+6q_{0}^{4}\epsilon+1\right)X_{\pm},\\
Q_{\pm}&=\lim_{x\rightarrow\pm\infty} U_{0}=\left(\begin{array}{cc}
0 & q_{\pm}\\
-q_{\pm}^{*} & 0
\end{array} \right).
\end{aligned}  \right.
\end{align}
Obviously, the two matrices $X_{\pm}$ and  $T_{\pm}$ are commutative, i.e., $[X_{\pm},T_{\pm}]=0$, which means that we can find a reversible matrix $E_{\pm}$ to diagonalize the two matrices. Since matrix $X_{\pm}$ has two eigenvalues $i\lambda$ and $-i\lambda$, it follows that two eigenvalues of $T_{\pm}$ can be known from
Eq. \eqref{Q6}, and further we can get
\begin{align}\label{Q7}
&\left\{\begin{aligned}
X_{\pm}E_{\pm}&=-i\lambda E_{\pm}\sigma_{3},\\
T_{\pm}E_{\pm}&=-i\lambda E_{\pm}\left(16k^{4}\epsilon-8k^{2}q_{0}^{2}\epsilon
+2k+6q_{0}^{4}\epsilon+1\right)\sigma_{3},
\end{aligned} \right.
\end{align}
with
\begin{align}\label{Q8}
&\left\{\begin{aligned}
\lambda&=\sqrt{k^{2}+q_{0}^{2}},\\
E_{\pm}&=\left(\begin{array}{cc}
1 & -\frac{iq_{\pm}}{k+\lambda}\\
-\frac{iq_{\pm}^{*}}{k+\lambda} & 1
\end{array}\right)=\mathbb{I}-\frac{i}{k+\lambda}\sigma_{3}Q_{\pm}.
\end{aligned}\right.
\end{align}
Obviously,
\begin{align}\label{bu-1}
\det E_{\pm}=1+\frac{q_{0}^{2}}{(k+\lambda)^{2}}\triangleq\gamma,\quad
E_{\pm}^{-1}=\frac{1}{\gamma}\left(\mathbb{I}+\frac{i}{k+\lambda}
\sigma_{3}Q_{\pm}\right).
\end{align}
\subsection{Riemann surface and uniformization coordinate}
The eigenvalue of the matrix $X_{\pm}$ reads $\pm i\lambda=\pm i\sqrt{k^{2}+q_{0}^{2}}$, which are branched functions with  multiple values. To avoid this situation, we introduce a two-sheeted Riemann surface defined by
\begin{align}\label{Q9}
\lambda^{2}=k^{2}+q_{0}^{2}=(k+iq_{0})(k-iq_{0}),
\end{align}
here the two-sheeted Riemann surface obtained by gluing two complex $k$-plane $S_{1}$ and $S_{2}$ along the segment $[-iq_{0}, iq_{0}]$. Obviously Eq. \eqref{Q9}
 can be written in polar coordinates, i.e., $k+iq_{0}=r_{1}e^{i\theta_{1}}, k-iq_{0}=r_{2}e^{i\theta_{2}}, -\frac{\pi}{2}<\theta_{1},\theta_{2}<\frac{3\pi}{2},$
which determined two single value functions
\begin{align}\label{Q11}
\lambda(k)=&\left\{\begin{aligned}
&(r_{1}r_{2})^\frac{1}{2}e^\frac{{\theta_{1}+\theta_{2}}}{2}, \quad &on\quad S_{1},\\
-&(r_{1}r_{2})^\frac{1}{2}e^\frac{{\theta_{1}+\theta_{2}}}{2}, \quad &on\quad S_{2}.
\end{aligned} \right.
\end{align}

To avoid complexity of Riemann surface, Shabat pointed that it is possible to introduce a affine parameter $z$ \cite{Faddeev-1987}, that is
\begin{align}\label{Q12}
z=k+\lambda,
\end{align}
further we get two single value functions
\begin{align}\label{Q13}
k(z)=\frac{1}{2}\left(z-\frac{q_{0}^{2}}{z}\right),\quad \lambda(z)=\frac{1}{2}\left(z+\frac{q_{0}^{2}}{z}\right).
\end{align}
From the relationships   \eqref{Q11} and   \eqref{Q13}, we have the following summary
\begin{table}[htbp]
\centering
\caption{Relationship between different planes}
\begin{tabular}{cccccccc}
\hline
&$S_1$ of $k$-plane & $+$ & $S_{2}$ of $k$-plane &\ $\rightarrow$ &\ $\lambda$-plane &\ $\rightarrow$ &\ $z$-plane \\ \hline
&$Imk>0$ & $+$ & $Imk<0$ & $\rightarrow$ & $Im\lambda>0$ & $\rightarrow$ & $D^{+}=\left\{z| (|z|^{2}-q_{0}^{2})Imz>0\right\}$ \\ \hline
&$Imk<0$ & $+$ & $Imk>0$ & $\rightarrow$ & $Im\lambda<0$ & $\rightarrow$ & $D^{-}=\left\{z| (|z|^{2}-q_{0}^{2})Imz<0\right\}$ \\ \hline
&$[-iq_{0},iq_{0}]$ & $+$ & $[-iq_{0},iq_{0}]$ & $\rightarrow$ & $[-q_{0},q_{0}]$ & $\rightarrow$ & $|z|^{2}-q_{0}^{2}=0$ \\ \hline
\end{tabular}
\end{table}

\centerline{\begin{tikzpicture}[scale=0.5]
\path [fill=yellow] (1,-5) -- (9,-5) to (9,-9) -- (1,-9);
\filldraw[pink, line width=0.5](-1,-5)--(3,-5) arc (-180:0:2);
\path [fill=pink] (1,-1) -- (9,-1) to (9,-5) -- (1,-5);
\filldraw[yellow, line width=0.5](3,-5)--(7,-5) arc (0:180:2);
\path [fill=pink] (-9,5)--(-9,9) to (-1,9) -- (-1,5);
\path [fill=yellow] (-9,1)--(-9,5) to (-1,5) -- (-1,1);
\path [fill=pink] (1,1)--(1,5) to (9,5) -- (9,1);
\path [fill=yellow] (1,5)--(1,9) to (9,9) -- (9,5);
\path [fill=pink] (-1,-1)--(-1,-5) to (-9,-5) -- (-9,-1);
\path [fill=yellow] (-1,-5)--(-1,-9) to (-9,-9) -- (-9,-5);
\filldraw[red, line width=0.5] (2,2) to (-2,-2)[->];
\filldraw[red, line width=0.5] (-2,-8) to (2,-8)[->];
\draw[fill] (-5,5)node[below]{} circle [radius=0.035];
\draw[fill] (5,5)node[below]{} circle [radius=0.035];
\draw[fill] (-5,-5)node[below]{} circle [radius=0.035];
\draw[fill] (5,-5)node[below]{} circle [radius=0.035];
\draw[-][thick](-9,5)--(-8,5);
\draw[-][thick](-8,5)--(-7,5);
\draw[-][thick](-7,5)--(-6,5);
\draw[-][thick](-6,5)--(-5,5);
\draw[-][thick](-5,5)--(-4,5);
\draw[-][thick](-4,5)--(-3,5);
\draw[-][thick](-3,5)--(-2,5);
\draw[-][thick](-2,5)--(-1,5)[->][thick]node[above]{$Rek$};;
\draw[-][thick](-5,1)--(-5,2);
\draw[-][thick](-5,2)--(-5,3);
\draw[-][thick](-5,3)--(-5,4);
\draw[-][thick](-5,4)--(-5,5);
\draw[-][thick](-5,5)--(-5,6);
\draw[-][thick](-5,6)--(-5,7);
\draw[-][thick](-5,7)--(-5,8);
\draw[-][thick](-5,8)--(-5,9)[->] [thick]node[above]{$Imk$};
\draw[-][thick](1,5)--(2,5);
\draw[-][thick](2,5)--(3,5);
\draw[-][thick](3,5)--(4,5);
\draw[-][thick](4,5)--(5,5);
\draw[-][thick](5,5)--(6,5);
\draw[-][thick](6,5)--(7,5);
\draw[-][thick](7,5)--(8,5);
\draw[-][thick](8,5)--(9,5)[->][thick]node[above]{$Rek$};
\draw[-][thick](5,1)--(5,2);
\draw[-][thick](5,2)--(5,3);
\draw[-][thick](5,3)--(5,4);
\draw[-][thick](5,4)--(5,5);
\draw[-][thick](5,5)--(5,6);
\draw[-][thick](5,6)--(5,7);
\draw[-][thick](5,7)--(5,8);
\draw[-][thick](5,8)--(5,9);
\draw[-][thick](-9,-5)--(-8,-5);
\draw[-][thick](-8,-5)--(-7,-5);
\draw[-][thick](-7,-5)--(-6,-5);
\draw[-][thick](-6,-5)--(-5,-5);
\draw[-][thick](-5,-5)--(-4,-5);
\draw[-][thick](-4,-5)--(-3,-5);
\draw[-][thick](-3,-5)--(-2,-5);
\draw[-][thick](-2,-5)--(-1,-5)[->][thick]node[above]{$Re\lambda$};
\draw[-][thick](-5,-1)--(-5,-2);
\draw[-][thick](-5,-2)--(-5,-3);
\draw[-][thick](-5,-3)--(-5,-4);
\draw[-][thick](-5,-4)--(-5,-5);
\draw[-][thick](-5,-5)--(-5,-6);
\draw[-][thick](-5,-6)--(-5,-7);
\draw[-][thick](-5,-7)--(-5,-8);
\draw[-][thick](-5,-8)--(-5,-9);
\draw[-][thick](1,-5)--(2,-5);
\draw[-][thick](2,-5)--(3,-5);
\draw[-][thick](3,-5)--(4,-5);
\draw[-][thick](4,-5)--(5,-5);
\draw[-][thick](5,-5)--(6,-5);
\draw[-][thick](6,-5)--(7,-5);
\draw[-][thick](7,-5)--(8,-5);
\draw[-][thick](8,-5)--(9,-5)[->][thick]node[above]{$Rez$};
\draw[-][thick](5,-1)--(5,-2);
\draw[-][thick](5,-2)--(5,-3);
\draw[-][thick](5,-3)--(5,-4);
\draw[-][thick](5,-4)--(5,-5);
\draw[-][thick](5,-5)--(5,-6);
\draw[-][thick](5,-6)--(5,-7);
\draw[-][thick](5,-7)--(5,-8);
\draw[-][thick](5,-8)--(5,-9);
\draw[->](5,9)[thick]node[above]{$Imk$};
\draw[->](-5,-1)[thick]node[above]{$Im\lambda$};
\draw[->](5,-1)[thick]node[above]{$Imz$};
\draw[fill] (-5,7) circle [radius=0.055]node[left]{\footnotesize$iq_{0}$};
\draw[fill] (-5,3) circle [radius=0.055]node[left]{\footnotesize$-iq_{0}$};
\draw[fill] (5,7) circle [radius=0.055]node[left]{\footnotesize$iq_{0}$};
\draw[fill] (5,3) circle [radius=0.055]node[left]{\footnotesize$-iq_{0}$};
\draw[fill] (-7,-5) circle [radius=0.055]node[below]{\footnotesize$q_{0}$};
\draw[fill] (-3,-5) circle [radius=0.055]node[below]{\footnotesize$-q_{0}$};
\draw(5,-5) [red, line width=1] circle(2);
\filldraw[red, line width=1.5] (-5,7) to (-5,3);
\filldraw[red, line width=1.5] (5,7) to (5,3);
\filldraw[red, line width=1.5] (-7,-5) to (-3,-5);
\draw[fill][black] (-8,7) [thick]node[right]{\footnotesize$S_{1}$};
\draw[fill][black] (2,7) [thick]node[right]{\footnotesize$S_{2}$};
\draw[fill][black] (-4,7) [thick]node[right]{\footnotesize$Imk>0$};
\draw[fill][black] (-4,3) [thick]node[right]{\footnotesize$Imk<0$};
\draw[fill][black] (6,7) [thick]node[right]{\footnotesize$Imk>0$};
\draw[fill][black] (6,3) [thick]node[right]{\footnotesize$Imk<0$};
\draw[fill][black] (-4,-7) [thick]node[right]{\footnotesize$Im\lambda<0$};
\draw[fill][black] (-4,-3) [thick]node[right]{\footnotesize$Im\lambda>0$};
\draw[fill][black] (7,-7) [thick]node[right]{\footnotesize$D_{-}$};
\draw[fill][black] (7,-3) [thick]node[right]{\footnotesize$D_{+}$};
\draw[fill][red] (0,5) node[]{\footnotesize$+$};
\draw[fill][black] (-2,0) [thick]node[right]
{\footnotesize$\lambda=\sqrt{k^{2}+q_{0}^{2}}$};
\draw[fill][black] (0,-8) [thick]node[below]
{\footnotesize$\lambda=(z+q_{0}^{2}/z)/2$};
\end{tikzpicture}}
\noindent {\small \textbf{Figure 1.} Transformation relation from $k$ two-sheeted Riemann surface, $\lambda$-plane and $z$-plane.}

\subsection{Jost functions and scattering matrix and their analyticity}
All values $k$ satisfying $\lambda(k)\in\mathbb{R}$ on each sheet constitute the continuous spectrum $\Sigma_{k}$, namely, $\Sigma_{k}=\mathbb{R}\cup [-iq_{0},iq_{0}]$. After transformation, the corresponding set in the complex $z$-plane is $\Sigma_{k}=\mathbb{R}\cup C_{0}$, where $C_{0}$ is the circle of radius $q_{0}$.  The subscript will be omitted  next for simplicity, hereafter we discuss the direct and inverse problem in the complex $z$-plane instead of the complex two-sheeted Riemann surface. Due to the situation, the Jost functions $\phi_{\pm}=\phi_{\pm}(x,t;z)$ satisfying the lax pair Eq. \eqref{Q4} can be defined for all $z\in\Sigma$ from Eq. \eqref{Q7}
\begin{align}\label{Q14}
\phi_{\pm}\sim E_{\pm}(z)e^{-i\theta(x,t;z)\sigma_{3}}+o(1) \quad x\rightarrow \pm\infty,
\end{align}
with $\theta(x,t;z)=\lambda\left[x+\left(16k^{4}\epsilon-8k^{2}q_{0}^{2}\epsilon
+2k+6q_{0}^{4}\epsilon+1\right)t\right]$.
Introducing the modified Jost functions
\begin{align}\label{Q16}
u_{\pm}(x,t;z)=\phi_{\pm}(x,t;z)e^{i\theta(x,t;z)\sigma_{3}}\rightarrow E_{\pm},
\quad x\rightarrow\pm\infty,
\end{align}
we get the solutions $u_{\pm}(x,t;z)$
\begin{align}\label{Q17}
\begin{matrix}
u_{-}(x,t;z)=Y_{-}+\int_{-\infty}^{x}Y_{-}e^{-i\lambda(x-y)\sigma_{3}}Y_{-}^{-1}\Delta Q_{-}(y,t)u_{-}(y,t;z)e^{i\lambda(x-y)\sigma_{3}}\, dy,\\
u_{+}(x,t;z)=Y_{+}-\int_{x}^{\infty}Y_{+}e^{-i\lambda(x-y)\sigma_{3}}Y_{+}^{-1}\Delta Q_{+}(y,t)u_{+}(y,t;z)e^{i\lambda(x-y)\sigma_{3}}\, dy,
\end{matrix}
\end{align}
via the Volterra integral equations.
By using the above integral equations, the Theorem 2.1 can be obtained, which is proved in Appendix A.

\begin{thm}
If $q(x,t)-q_{-}\in L^{1}(-\infty,a)$, or $q(x,t)-q_{+}\in L^{1}(a,\infty)$ with all constant $a\in\mathbb{R}$, the modify Jost functions $u_{\pm}(x,t;z)$ can be analytically extended onto the corresponding regions of the $z$-plane, that is
\begin{align}\label{Q18}
&\left\{\begin{aligned}
u_{-,1}(x,t;z), u_{+,2}(x,;z) \in D^{+},\\
u_{-,2}(x,t;z), u_{+,1}(x,;z) \in D^{-}.
\end{aligned}\right.
\end{align}
\end{thm}

Next, we will introduce the scattering matrix. It is obvious that the trace of the solution $\phi$ satisfying the Lax pair Eq. \eqref{Q4} is zero, that is $trU=trV=0$, which implies $\partial_{x}(\det\phi)=\partial_{t}(\det\phi)=0$ from \cite{Liu}. Further, we have
\begin{align}\label{Q19}
\det\phi_{\pm}(x,t;z)=\det E_{\pm}(z)=\gamma(z)\quad x,t\in\mathbb{R}, \quad z\in\Sigma.
\end{align}
Denoting $\Sigma_{1}=\Sigma\setminus \{\pm iq_{0}\}$, for any $z\in\Sigma_{1}$ we know that $\phi_{\pm}$ are two fundamental solutions of the scattering problem. There  exists a $2\times2$  matrix $S(z)$ (it's independent of the variable $x$ and $t$) such that
\begin{align}\label{Q20}
\phi_{+}(x,t;z)=\phi_{-}(x,t;z)S(z),\quad
S(z)=\left(\begin{array}{cc}
         s_{11}(z) & s_{12}(z) \\
         s_{21}(z) & s_{22}(z) \\
       \end{array}\right), \quad
 z\in\Sigma_{1},
\end{align}
because the scattering problem is a first order homogeneous ordinary differential equation. Obviously \eqref{Q20} implies
\begin{align}\label{Q21}
\begin{split}
\phi_{+,1}(x,t;z)&=s_{11}(z)\phi_{-,1}(x,t;z)+s_{21}(z)\phi_{-,2}(x,t;z),\\
\phi_{+,2}(x,t;z)&=s_{12}(z)\phi_{-,1}(x,t;z)+s_{22}(z)\phi_{-,2}(x,t;z),
\end{split}
\end{align}
where $\phi_{\pm,j}$ represent the column element of $\phi_{\pm}$.  In additional, the Eqs. \eqref{Q19} and \eqref{Q20} yield obviously $\det S(z)=1$.

In order to obtain the analyticity of scattering coefficient $(s_{ij})_{1}^{2}$, we define  the Wronskian determinant, i.e., $Wr(u,v)=u_{1}v_{2}-u_{2}v_{1}$ with $u=(u_{1},u_{2})^{T}$ and $v=(v_{1},v_{2})^{T}$, thus
\addtocounter{equation}{1}
\begin{align}
s_{11}(z)\gamma=Wr\left(\phi_{+,1},\phi_{-,2}\right), \quad
s_{22}(z)\gamma=Wr\left(\phi_{-,1},\phi_{+,2}\right),\tag{\theequation a} \label{s11}\\
s_{12}(z)\gamma=Wr\left(\phi_{+,2},\phi_{-,2}\right),\quad
s_{21}(z)\gamma=Wr\left(\phi_{-,1},\phi_{+,1}\right),\tag{\theequation b}
\label{s12}
\end{align}
which means $s_{11}(z)$ and $s_{22}(z)$ are analytic in $D^{-}$ and $D^{+}$, respectively, meanwhile $s_{12}(z)$ and $s_{21}(z)$ are not analytic, but continuous to $\Sigma_{1}$ from the Theorem 2.1. Also the reflection coefficients are introduced to construct RH problem,  namely
\begin{align}\label{Q22}
\rho(z)=s_{21}(z)/s_{11}(z),\quad \tilde{\rho}(z)=s_{12}(z)/s_{22}(z),\quad \forall z\in\Sigma.
\end{align}
\subsection{Symmetries}
By using the RH method to solve the initial value problem, we need to consider the symmetry of the potential function of the Lax pair, which can derive the symmetry of the eigenfunctions and further affect the symmetry of the scattering data, so that the distribution of spectral points can be obtained. It should be noted that the symmetry with non-zero boundary conditions is very complicated by the fact that $\lambda(k)$ will change the symbols from one sheet of the Riemann surface to the other, namely $\lambda_{2}(k)=-\lambda_{1}(k)$. Based on the affine parameter $z$,
one consider the following two kinds of symmetry, i.e., the first symmetry $z\mapsto z^{*}$ yields $(k,\lambda)\mapsto(k^{*},\lambda^{*})$, and the second symmetry  $z\mapsto -q_{0}^{2}/z$ yields $(k,\lambda)\mapsto(k,-\lambda)$. It follows from the above two symmetries that
\begin{align}\label{Q23}
&\left\{ \begin{aligned}
&\ \phi_{\pm}(x,t;z)=\sigma_{0}\phi_{\pm}^{*}(x,t;z^{*})\sigma_{0},\\
&\ \phi_{\pm}(x,t;z)=-\frac{i}{z}\phi_{\pm}^{*}(x,t;-q_{0}^{2}/z)\sigma_{3}Q_{+},
\end{aligned}\right.
\end{align}
note that $\theta^{*}(x,t;z^{*})=\theta(x,t;z)$, $k(-q_{0}^{2}/z)=k(z)$, $\theta(x,t;-q_{0}^{2}/z)=-\theta(x,t;z)$ and $\sigma_{0}Q^{*}\sigma_{0}=Q^{\dag}=-Q$ with $\sigma_{0}=\left(\begin{array}{cc}
0 & 1\\
-1 & 0\\
\end{array}\right)$, which can derive the symmetry of the scattering matrix $S(z)$ combined with \eqref{Q20}
\begin{align}\label{Q24}
&\left\{ \begin{aligned}
&\ S^{*}(z^{*})=-\sigma_{0}S(z)\sigma_{0},\\
&\ S(z)=(\sigma_{3}Q_{-})^{-1}S(-q_{0}^{2}/z)\sigma_{3}Q_{+}.
\end{aligned}\right.
\end{align}
In terms of the symmetry of Jost functions and scattering matrix, one can deduce the symmetry relation between scattering coefficient and reflection coefficient by direct calculation, which is ultimately related to the distribution of spectral points
\begin{align}\label{Q25}
&\left\{ \begin{aligned}
&\ s_{22}(z)=s^{*}_{11}(z^{*}), \quad s_{12}(z)=-s^{*}_{21}(z^{*}),\\
&\ s_{11}(z)=\frac{q_{+}^{*}}{q_{-}^{*}}s_{22}(-q_{0}^{2}/z),
s_{12}(z)=\frac{q_{+}}{q_{-}^{*}}s_{21}(-q_{0}^{2}/z),\\
&\ \rho(z)=-\tilde{\rho}^{*}(z^{*}),\quad
\rho(z)=(q^{*}_{-}/q_{-})\tilde{\rho}(-q_{0}^{2}/z).
\end{aligned}\right.
\end{align}
\subsection{Discrete spectrum and residue conditions}
The discrete spectrum of the scattering problem consist of all values $z\in\mathbb{C}\setminus\Sigma$ satisfying the scattering problem admits eigenfunctions in $L^{2}(\mathbb{R})$. These discrete spectrum are respectively the zeros of $s_{11}(z)$ and $s_{22}(z)$ in the corresponding analytical region, namely for $z\in D^{-}$ and  $z\in D^{+}$. Now assume that $s_{11}(z)$ has $N$ simple zeros $z_{j}$, i.e., $s_{11}(z_{j})=0$, but $s'_{11}(z_{j})\neq 0$, $j=1,2, \cdots, N$, in the region $D^{-}\cap \{z\in\mathbb{C}: Imz<0\}$, besides $z_{j}$ satisfies $|z_{j}|>q_{0}$ and $Imz_{j}>0$. Therefore the Eq. \eqref{Q25} implies
\begin{align}\label{Q26}
s_{11}(z_{j})=0\Longleftrightarrow s_{22}(z_{j}^{*})=0 \Longleftrightarrow s_{22}(-q_{0}^{2}/z_{j})=0\Longleftrightarrow
s_{11}(-q_{0}^{2}/z_{j}^{*})=0,
\end{align}
which means that a quartet of discrete eigenvalues are derived, i.e.,
\begin{align}\label{Q27}
\mathbb{Z}=\left\{z_{j}, -q_{0}^{2}/{z_{j}^{*}},
z_{j}^{*}, -q_{0}^{2}/{z_{j}}\right\}_{j=1}^{N}.
\end{align}

\centerline{\begin{tikzpicture}[scale=0.65]
\draw[fill] (0,-2)node[below]{} circle [radius=0.055];
\draw[fill] (0,2)node[below]{} circle [radius=0.055];
\draw[fill] (2,0)node[below]{} circle [radius=0.055];
\draw[fill] (-2,0)node[below]{} circle [radius=0.055];
\draw[fill] (0,0)node[below]{} circle [radius=0.055];
\path [fill=pink] (-4,0)--(-4,4) to (4,4) -- (4,0);
\filldraw[yellow, line width=0.5](-4,0)--(-2,0) arc (180:0:2);
\path [fill=yellow] (-4,0)--(-4,-4) to (4,-4) -- (4,0);
\filldraw[pink, line width=0.5](-4,0)--(-2,0) arc (-180:0:2);
\draw[->][thick](-4,0)--(-3,0);
\draw[-][thick](-3,0)--(-2,0)node[below right]{\footnotesize$0^{-}$};;
\draw[-][thick](-2,0)--(-1,0);
\draw[<-][thick](-1,0)--(0,0);
\draw[-][thick](0,0)--(1,0);
\draw[<-][thick](1,0)--(2,0)node[below right]{\footnotesize$0^{+}$};
\draw[->][thick](2,0)--(3,0);
\draw[->][thick](3,0)--(4,0)[thick]node[right]{$Rez$};
\draw[-][thick](0,0)--(0,1);
\draw[-][thick](0,1)--(0,2)node[below right]{\footnotesize$iq_{0}$};
\draw[-][thick](0,2)--(0,3);
\draw[->][thick](0,3)--(0,4)[thick]node[right]{$Imz$};
\draw[-][thick](0,0)--(0,-1);
\draw[-][thick](0,-1)--(0,-2)node[below right]{\footnotesize$-iq_{0}$};
\draw[-][thick](0,-2)--(0,-3);
\draw[-][thick](0,-3)--(0,-4);
\draw[->][red, line width=0.8] (2,0) arc(0:220:2);
\draw[->][red, line width=0.8] (2,0) arc(0:330:2);
\draw[->][red, line width=0.8] (2,0) arc(0:-330:2);
\draw[->][red, line width=0.8] (2,0) arc(0:-220:2);
\draw[fill][blue] (2.5,2.5) circle [radius=0.035][thick]node[right]{\footnotesize$z_{j}$};
\draw[fill] (2.5,-2.5) circle [radius=0.035][thick]node[right]{\footnotesize$z_{j}^{*}$};
\draw[fill] (-1,1) circle [radius=0.035][thick]node[right]{\footnotesize$-\frac{q_{0}^{2}}{z_{j}}$};
\draw[fill][blue] (-1,-1) circle [radius=0.035][thick]node[right]{\footnotesize$-\frac{q_{0}^{2}}{z_{j}^{*}}$};
\end{tikzpicture}}

\noindent {\small \textbf{Figure 2.} The distribution of the set of discrete spectral points on the $z$-plane,
where the blue spectral points are the zeros of $s_{11}(z)$, the black spectral points are the zeros of $s_{22}(z)$, and the red contour is the jump condition about the RH problem Eq. \eqref{Jump}. }

In what follows, we derive the residue conditions of the Jost functions at discrete spectral points, which will be needed in the process of inverse scattering. The expression of Eq.  \eqref{s11} can be written equivalently as
\begin{align}\label{Q28}
\phi_{+,1}(z_{n})=b_{-}(z_{n})\phi_{-,2}(z_{n}),
\end{align}
for $z_{n}\in\mathbb{Z}\cap D^{-}$ is the simple zero of $s_{11}(z)$, where $b_{-}(z_{n})$ is a constant. The residue condition then can be written as
\begin{align}\label{Q29}
\mathop{Res}_{z=z_{n}}\left[\frac{\phi_{+,1}(z)}{s_{11}(z)}\right]=
\frac{\phi_{-,1}(z_{n})}{s'_{11}(z_{n})}=\frac{b_{+}(z_{n})}
{s'_{11}(z_{n})}\phi_{-,2}(z_{n}).
\end{align}
Similarly from the another expression of Eq. \eqref{s11}, we can derive the following residue condition as $z_{n}^{*}\in\mathbb{Z}\cap D^{+}$ is the simple zero of $s_{22}(z)$
\begin{align}\label{Q30}
\mathop{Res}_{z=z_{n}^{*}}\left[\frac{\phi_{+,2}(z)}{s_{22}(z)}\right]=
\frac{\phi_{+,2}(z_{n}^{*})}{s'_{22}(z_{n}^{*})}=
\frac{b_{+}(z_{n}^{*})}{s'_{22}(z_{n}^{*})}\phi_{-,1}(z_{n}^{*}),
\end{align}
here $b_{+}(z_{n}^{*})$ is a constant parameter. Introducing the notations
\begin{align}\label{Q31}
C_{-}[z_{n}]=\frac{b_{-}(z_{n})}{s'_{11}(z_{n})},\quad
z_{n}\in\mathbb{Z}\cap D^{-},\quad
C_{+}[z_{n}^{*}]=\frac{b_{+}(z_{n}^{*})}{s'_{22}(z_{n}^{*})},\quad
z_{n}^{*}\in\mathbb{Z}\cap D^{+},
\end{align}
and combining the symmetries of Eq. \eqref{Q25}, one has the results
\begin{align}\label{Q32}
\left\{\begin{aligned}
C_{-}[z_{n}]&=-C_{+}^{*}[z_{n}^{*}],\quad C_{-}[z_{n}]=\frac{z_{n}^{2}}{q_{-}^{2}}C_{+}
\left[-\frac{q_{0}^{2}}{z_{n}}\right],\quad z_{n}\in\mathbb{Z}\cap D^{-},\\
C_{-}[z_{n}]&=-C_{+}^{*}[z_{n}^{*}]=\frac{z_{n}^{2}}{q_{-}^{2}}
C_{+}\left[-\frac{q_{0}^{2}}{z_{n}}\right]=-\frac{z_{n}^{2}}{q_{-}^{2}}
C_{-}^{*}\left[-\frac{q_{0}^{2}}{z_{n}^{*}}\right],\quad z_{n}\in\mathbb{Z}\cap D^{-}.
\end{aligned}\right.
\end{align}
\section{Inverse scattering problem with simple zeros}
In the inverse problem, we first recover the modified eigenfunctions from the scattering data, and then recover the potential according to the asymptotic behavior of the Jost functions in the spectral parameters.
\subsection{Riemann-Hilbert problem}

To construct the RH \cite{Shabat-1976} problem, it is necessary to seed two functions that are analytic in the appropriate region. In terms of the previous analysis, resorting to Eqs. \eqref{Q16} and \eqref{Q20} we obtain the following two sectionally analytic functions
\begin{align}\label{Q33}
\left\{\begin{aligned}
\frac{u_{+,1}(z)}{s_{11}(z)}&=u_{-,1}(z)+\frac{s_{21}(z)}{s_{11}(z)}
e^{2i\theta(z)}u_{-,2}(z),\\
\frac{u_{+,2}(z)}{s_{22}(z)}&=\frac{s_{12}(z)}{s_{22}(z)}
e^{-2i\theta(z)}u_{-,1}(z)+u_{-,2}(z).
\end{aligned}\right.
\end{align}
By using the modified eigenfunctions, the matrix $M(x,t;z)$ can further be defined as
\begin{align}\label{Matr}
M(x,t;z)=\left\{\begin{aligned}
&M^{+}(x,t;z)=\left(u_{-,1}(x,t;z),\frac{u_{+,2}(x,t;z)}{s_{22}(z)}\right), \quad z\in D^{+},\\
&M^{-}(x,t;z)=\left(\frac{u_{+,1}(x,t;z)}{s_{11}(z)},u_{-,2}(x,t;z)\right), \quad z\in D^{-}.
\end{aligned}\right.
\end{align}
The asymptotic conditions of the Jost functions and scattering matrix are needed now, similarly \cite{NZBC-14}, one has
\begin{align}\label{Q34}
u_{\pm}(x,t;z)=
\left\{\begin{aligned}
&\mathbb{I}+O(1/z),\quad\quad &z\rightarrow\infty,\\
&-\frac{i}{z}\sigma_{3}Q_{\pm}+O(1),\quad &z\rightarrow0.
\end{aligned}
\right.
\end{align}
 From Eqs. \eqref{s11} and \eqref{s12}, the asymptotic conditions about scattering matrix can further derived
\begin{align}\label{Sz}
S(z)=\left\{\begin{aligned}
\mathbb{I}+O(1/z),  \qquad \quad\qquad\qquad z\rightarrow\infty,\\
diag(q_{-}/q_{+},q_{+}/q_{-})+O(z),\quad z\rightarrow0.
\end{aligned}
\right.
\end{align}
Using the above results, we give the following theorem
\begin{thm} A matrix RH probelm:\\
$\bullet$ Analyticity:  $M(x,t;z)$ is meromorphic in $C\setminus\Sigma$.\\
$\bullet$ Jump condition: \begin{align}\label{Jump}
M^{+}(x,t;z)=M^{-}(x,t;z)(\mathbb{I}-G(x,t;z)),\quad z\in\Sigma,
\end{align}
with the jump matrix is
\begin{align*}
G(x,t;z)=e^{i\theta(z)\hat{\sigma}_{3}}\left(
\begin{array}{ccc}
  0 & -\tilde{\rho}(z) \\
  \rho(z) & \rho(z)\tilde{\rho}(z)
\end{array} \right). \notag
\end{align*}\\
$\bullet$
Asymptotic condition: \begin{align}\label{jianjin}
M^{\pm}(x,t;z)=\left\{
\begin{aligned}
\mathbb{I}+O(1/z), \qquad z\rightarrow\infty,\\
-(i/z)\sigma_{3}Q_{-}+O(1), \quad z\rightarrow0.
\end{aligned}
\right.
\end{align}
\end{thm}
The quartet of discrete eigenvalues $\mathbb{Z}$ is equivalent to the set $\mathbb{Z}=\left\{\eta_{n},\hat{\eta}_{n}\right\}$, $(n=1,2,\cdots,N)$, as $\eta_{n}=z_{n}$ and $\hat{\eta}_{n}=-q_{0}^{2}/z_{n-N}^{*}$ for $(n=1,2,\cdots,N)$ and $(n=N+1,N+2,\cdots,2N)$, respectively, as well as $\hat{\eta}_{n}=-q_{0}^{2}/\eta_{n}$.
Then the RH problem Eq. \eqref{Jump} can be solved with projection projectors and Plemelj's formulae
\begin{align}\label{Q35}
\begin{split}
M(x,t;z)=&\mathbb{I}-\frac{i}{z}\sigma_{3}Q_{+}+\sum_{n=1}^{2N}\frac
{\mathop{Res}\limits_{z=\hat{\eta}_{n}}M^{+}(z)}{z-\hat{\eta}_{n}}+\sum_{n=1}^{2N}\frac
{\mathop{Res}\limits_{z=\eta_{n}}M^{-}(z)}{z-\eta_{n}}\\
&+\frac{1}{2\pi i}\int_{\Sigma}\frac{M(x,t;s)^{-}G(x,t;s)}{s-z}\,ds,\quad
z\in\mathbb{C}\setminus\Sigma,
\end{split}
\end{align}
where the $\int_{\Sigma}$ implies the contour shown in Fig. 2, and the projection projectors are
\begin{align}\label{Cauchy}
P_{\pm}[f](z)=\frac{1}{2\pi i}\int_{\Sigma}\frac{f(\zeta)}{\zeta-(z\pm i0)}\,d\zeta,
\end{align}
here the $\int_{\Sigma}$ implies the integral along the oriented contour shown in Fig. 2. The notation $z\pm i0$ mean that the limit is taken from the left/right of $z$ $ (z\in\Sigma)$ respectively.

Then we can get a closed algebraic system by using the residual condition of the modified eigenfunctions $M^{\pm}$ at the discrete spectrum point $\mathbb{Z}$, and finally construct the exact expression of the potential $q$ by combining the asymptotic behavior of the modified eigenfunction when the spectrum parameter tends to infinity. Resorting to Eqs. \eqref{Q16}, \eqref{Q29} \eqref{Q30} and \eqref{Matr}, one has
\begin{align}\label{Q36}
\left\{\begin{aligned}
\mathop{Res}_{z=\hat{\eta}_{n}}M^{+}&=(0,C_{+}[\hat{\eta}_{n}]
e^{-2i\theta(\hat{\eta}_{n})}u_{-,1}(\hat{\eta}_{n})),\quad n=1,2,\cdots,2N,\notag \\
\mathop{Res}_{z=\eta_{n}}M^{-}&=(C_{-}[\eta_{n}]e^{2i\theta(\eta_{n})}u_{-,2}
(\eta_{n}),0), \quad n=1,2,\cdots,2N.
\end{aligned}\right.
\end{align}
Recalling Eq. \eqref{Matr}, one can calculate the part of Eq. \eqref{Q35}
\begin{align}
\frac{\mathop{Res}\limits_{z=\hat{\eta}_{n}}M^{+}(z)}{z-\hat{\eta}_{n}}+
\frac{\mathop{Res}\limits_{z=\eta_{n}}M^{-}(z)}{z-\eta_{n}}=\left(
\frac{C_{-}[\eta_{n}]e^{2i\theta(\eta_{n})}}{z-\eta_{n}}
u_{-,2}(\eta_{n}),
\frac{C_{+}[\hat{\eta}_{n}]e^{-2i\theta(\hat{\eta}_{n})}}{z-\hat{\eta}_{n}}
u_{-,1}(\hat{\eta}_{n})\right),
\end{align}
which implies from  Eq. \eqref{Q35}
\begin{align}\label{Q37}
\begin{split}
u_{-,2}(x,t;\eta_{s})=\left(\begin{array}{cc}
                       -iq_{-}/\eta_{s} \\
                        1
                     \end{array}\right)&+\sum_{n=1}^{2N}
\frac{C_{+}[\hat{\eta}_{n}]e^{-2i\theta(x,t;\hat{\eta}_{n})}}{\eta_{s}-\hat{\eta}_{n}}
u_{-,1}(x,t;\hat{\eta}_{n})\\
&+\frac{1}{2\pi i}\int_{\Sigma}\frac{(M^{-}G)_{2}(x,t;\zeta)}{\zeta-\eta_{s}}\,d\zeta,
\quad s=1, 2, \cdots, 2N.
\end{split}
\end{align}
From the symmetry Eq. \eqref{Q23}, one has
\begin{align}\label{Q38}
u_{-,2}(\eta_{s})=-\frac{iq_{-}}{\hat{\eta}_{s}}u_{-,1}(\hat{\eta}_{n}),\quad s=1, 2, \cdots, 2N,
\end{align}
further
\begin{align}\label{Q39}
\sum_{n=1}^{2N}\left(\frac{C_{+}[\hat{\eta}_{n}]e^{-2i\theta(\hat{\eta}_{n})}}
{\eta_{s}-\hat{\eta}_{n}}+\frac{iq_{-}}{\eta_{s}}\delta_{sn}\right)
u_{-,1}(\hat{\eta}_{n})+\left(\begin{array}{cc}
                      -\frac{iq_{-}}{\eta_{s}}  \\
                        1
                     \end{array}\right)+
  \frac{1}{2\pi i}\int_{\Sigma}\frac{(M^{-}G)_{2}(\eta)}{\eta-\eta_{s}}\,d\eta=0,
\end{align}
with the function $\delta_{sn}=1$ as $s=n$, but $\delta_{sn}=0$ as $s\neq n$, which forms a system of $2N$ equations with $2N$ unknowns $u_{-,1}(\hat{\eta}_{n})$ for $n=1,2,\cdots, 2N$. The expression of Eq. \eqref{Q39} combining with Eqs. \eqref{Q35} and \eqref{Q38} compose a closed system about the function $M(x,t;z)$ based on the scattering data. By using the asymptotic behaviour of $M(x,t;z)$ and the solution to the scattering problem Eq. \eqref{Q4}, namely $M(z)e^{i\theta(z)\sigma_{3}}$ , we can get the exact solution to the ifoNLS with simple zeros via comparing the coefficient of $z^{0}$.
\begin{prop}
The exact formal solution to ifoNLS with simple zeros
\begin{align}\label{Q40}
q(x,t)=-q_{-}-i\sum_{n=1}^{2N}C_{+}[\hat{\eta}_{n}]e^{-2i\theta(\hat{\eta}_{n})}
u_{-,11}(x,t;\hat{\eta}_{n})+\frac{1}{2\pi}\int_{\Sigma}(M^{-}G)_{12}(x,t;\eta)\,d\eta,
\end{align}
where the function $u_{-,1,1}$ is defined by from Eq. \eqref{Q39}
\begin{align}\label{u11}
\sum_{n=1}^{2N}\left(\frac{C_{+}[\hat{\eta}_{n}]e^{-2i\theta(\hat{\eta}_{n})}}
{\eta_{s}-\hat{\eta}_{n}}+\frac{iq_{-}}{\eta_{s}}\delta_{sn}\right)
u_{-,1,1}(\hat{\eta}_{n})- \frac{iq_{-}}{\eta_{s}} +
  \frac{1}{2\pi i}\int_{\Sigma}\frac{(M^{-}G)_{1,2}(\eta)}{\eta-\eta_{s}}\,d\eta=0.
  \end{align}
\end{prop}
In what follows,  we consider the solution with reflection-less potential, then Eqs. \eqref{Q40} and \eqref{u11} can be written as
\begin{align}
q(x,t)=-q_{-}-i\sum_{n=1}^{2N}C_{+}[\hat{\eta}_{n}]e^{-2i\theta(\hat{\eta}_{n})}
u_{-,1,1}(x,t;\hat{\eta}_{n}),\label{Q41}\\
\sum_{n=1}^{2N}\left(\frac{C_{+}[\hat{\eta}_{n}]e^{-2i\theta(\hat{\eta}_{n})}}
{\eta_{s}-\hat{\eta}_{n}}+\frac{iq_{-}}{\eta_{s}}\delta_{sn}\right)
u_{-,1,1}(x,t;\hat{\eta}_{n})- \frac{iq_{-}}{\eta_{s}}=0.
\end{align}
Letting
\begin{align}\label{xingshi}
\begin{split}
\mathcal {G}=(g_{sj})_{(2N)\times(2N)}, \varpi=(w_{j})_{(2N)\times 1},
\varrho=(v_{j})_{(2N)\times 1},\\
g_{sj}=\frac{w_{j}}{\eta_{s}-\hat{\eta}_{j}}+v_{s}\delta_{sj},
w_{j}=C_{+}[\hat{\eta}_{j}]e^{-2i\theta(\hat{\eta_{j}})},
v_{j}=\frac{iq_{-}}{\eta_{j}},
\end{split}
\end{align}
one can convert the solution Eq. \eqref{Q41} into the Theorem 1.1 by direct calculation of matrix properties.
\subsection{Soliton solutions}
In this section, we will use the exact expression of the solution Eq. \eqref{Jie-Q} to the Eq. \eqref{Q1} combined with specific appropriate parameters to describe the propagation behavior of the solution of the ifoNLS equation.\\
\noindent{\textbf{Case A:} For $N=1$, $z_{1}=3i/2$ and $\epsilon=0.01$, we get the breather wave solution, and when the initial value $q_{0}$ tends to zero, the periodic behavior of the solution gradually shifts upward, and the final solution tends to a bright soliton solution, which is similar to the solution constructed under the condition of zero boundary value. This phenomenon is shown in Figure 3, where $q_{0}=1, 0.5, 0.2, 0.01$ in figures (a-d). }\\
\noindent{\textbf{Case B:} For $N=1$, $q_{0}=1$ and $\epsilon=0.01$, the dynamic behavior of the solution with the discrete spectrum $z_{1}=\sqrt{2}e^{\frac{\pi i}{4}}$ and  $z_{1}=\sqrt{2}e^{\frac{\pi i}{6}}$ Eq. \eqref{Jie-Q} will be shown in Fig. 4. These discrete spectrum mean that the asymptotic phase difference are  $\pi$ and $4\pi/3$.}\\
\noindent{\textbf{Case C:} Figure 5 exhibits the interaction of two breather solitons for $N=2$, $q_{0}=1$, $\epsilon=0.01$, $z_{1}=1+\frac{3i}{2}$ and $z_{1}=-1+\frac{3i}{2}$. Similarly as $q_{0}\rightarrow 0$, the breather-breather wave solution tend to the bright-bright soliton solution.}

{\rotatebox{0}{\includegraphics[width=3.75cm,height=3.5cm,angle=0]{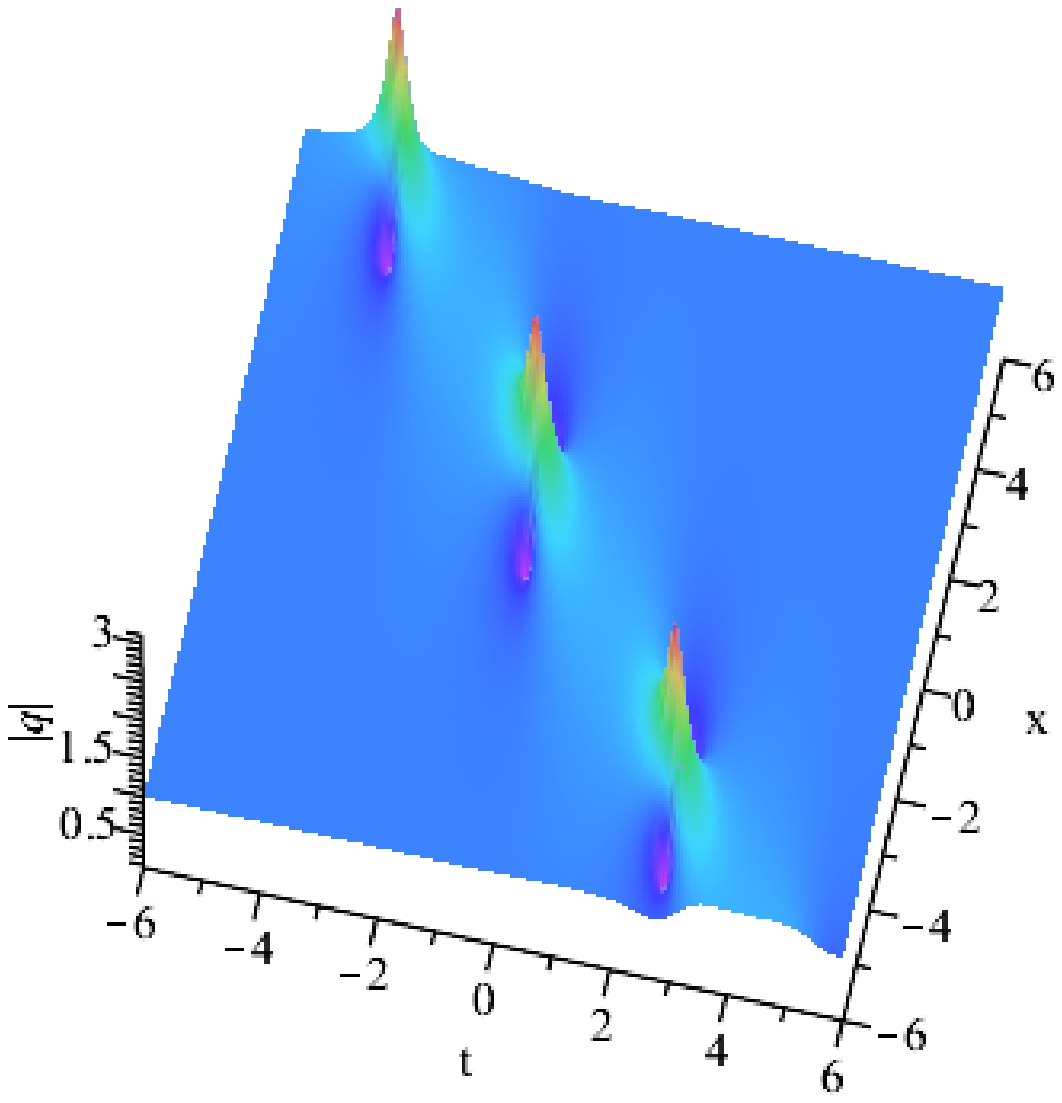}}}
\qquad\qquad\qquad\qquad
{\rotatebox{0}{\includegraphics[width=3.75cm,height=3.5cm,angle=0]{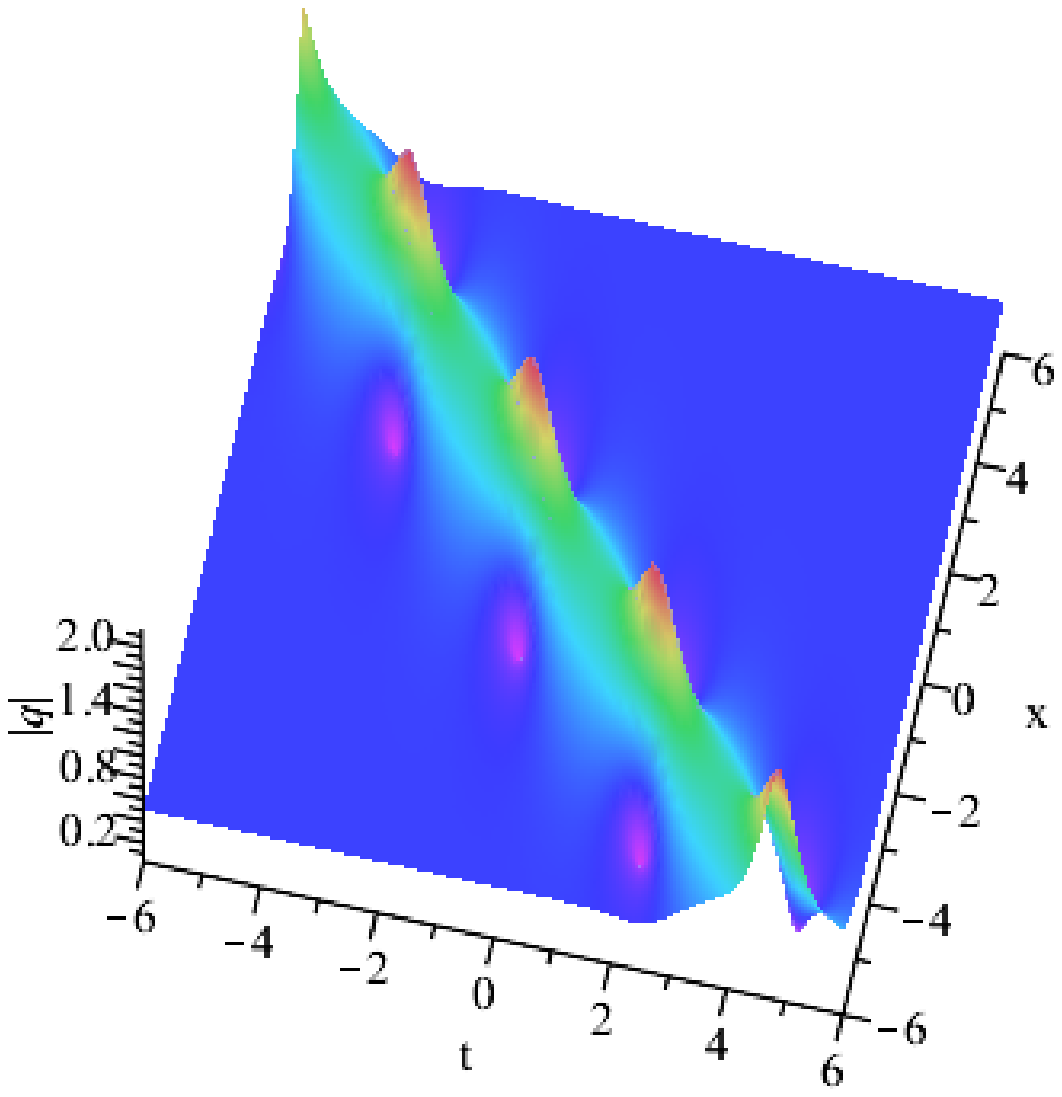}}}

 $\qquad\qquad(\textbf{a})\qquad \ \qquad\qquad\qquad\qquad\qquad \qquad\qquad\qquad(\textbf{b})$

{\rotatebox{0}{\includegraphics[width=3.75cm,height=3.5cm,angle=0]{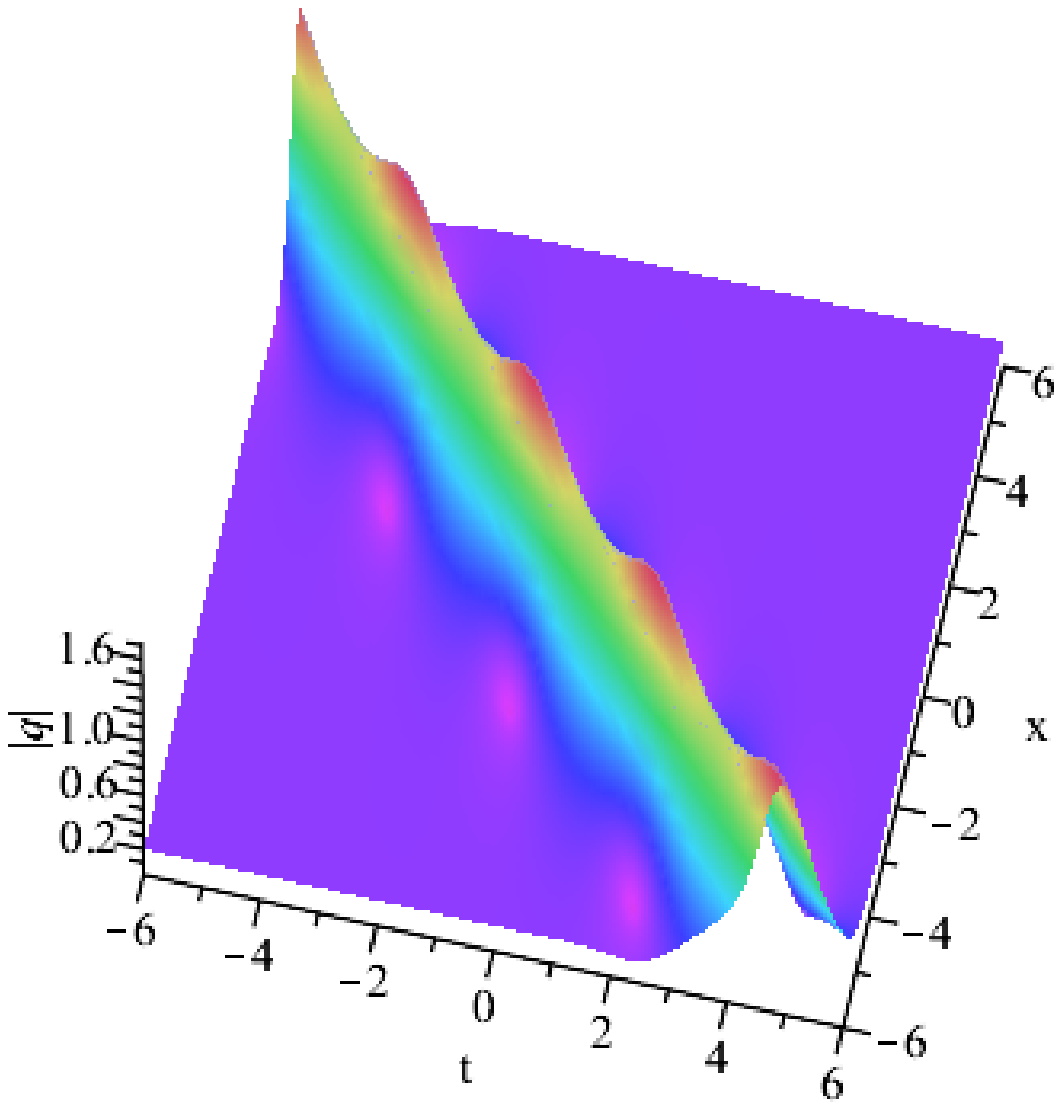}}}
\qquad\qquad\qquad\qquad
{\rotatebox{0}{\includegraphics[width=3.75cm,height=3.5cm,angle=0]{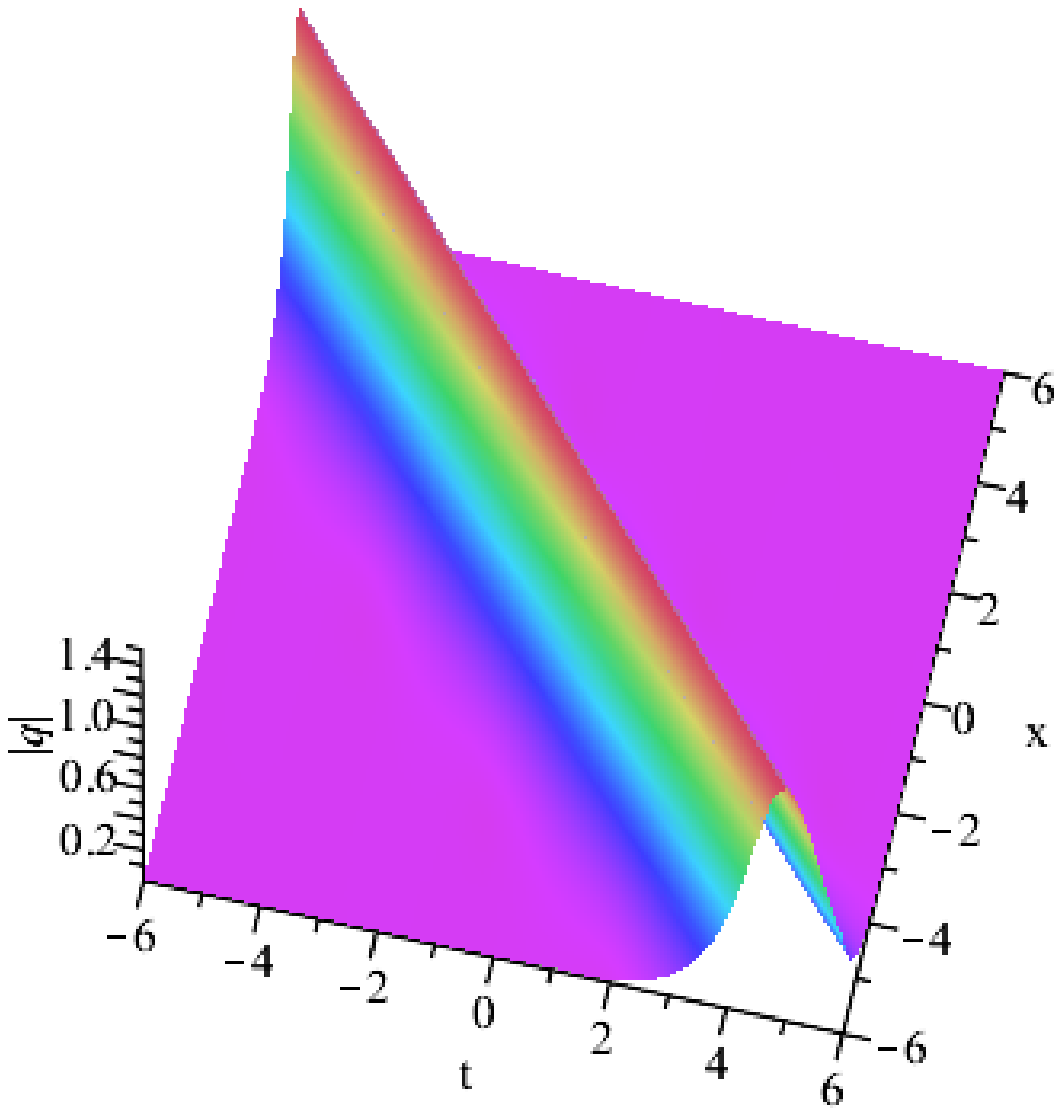}}}

 $\qquad\qquad(\textbf{c})\qquad \ \qquad\qquad\qquad\qquad\qquad \qquad\qquad\qquad(\textbf{d})$\\
\centerline{\noindent {\small \textbf{Figure 3.} Propagation of the solution Eq. \eqref{Jie-Q}.}}

{\rotatebox{0}{\includegraphics[width=3.75cm,height=3.5cm,angle=0]{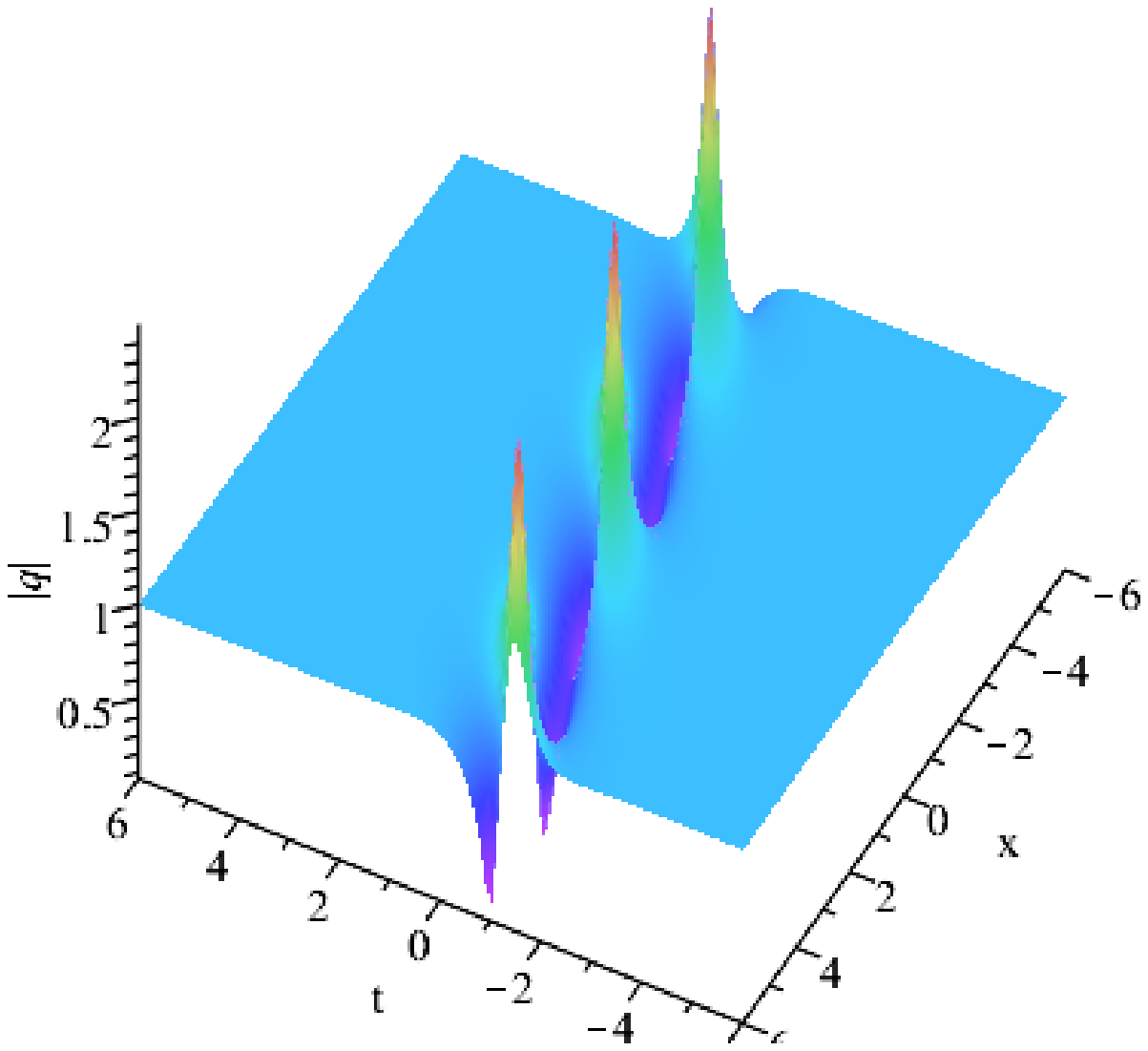}}}
\qquad\qquad\qquad\qquad
{\rotatebox{0}{\includegraphics[width=3.75cm,height=3.5cm,angle=0]{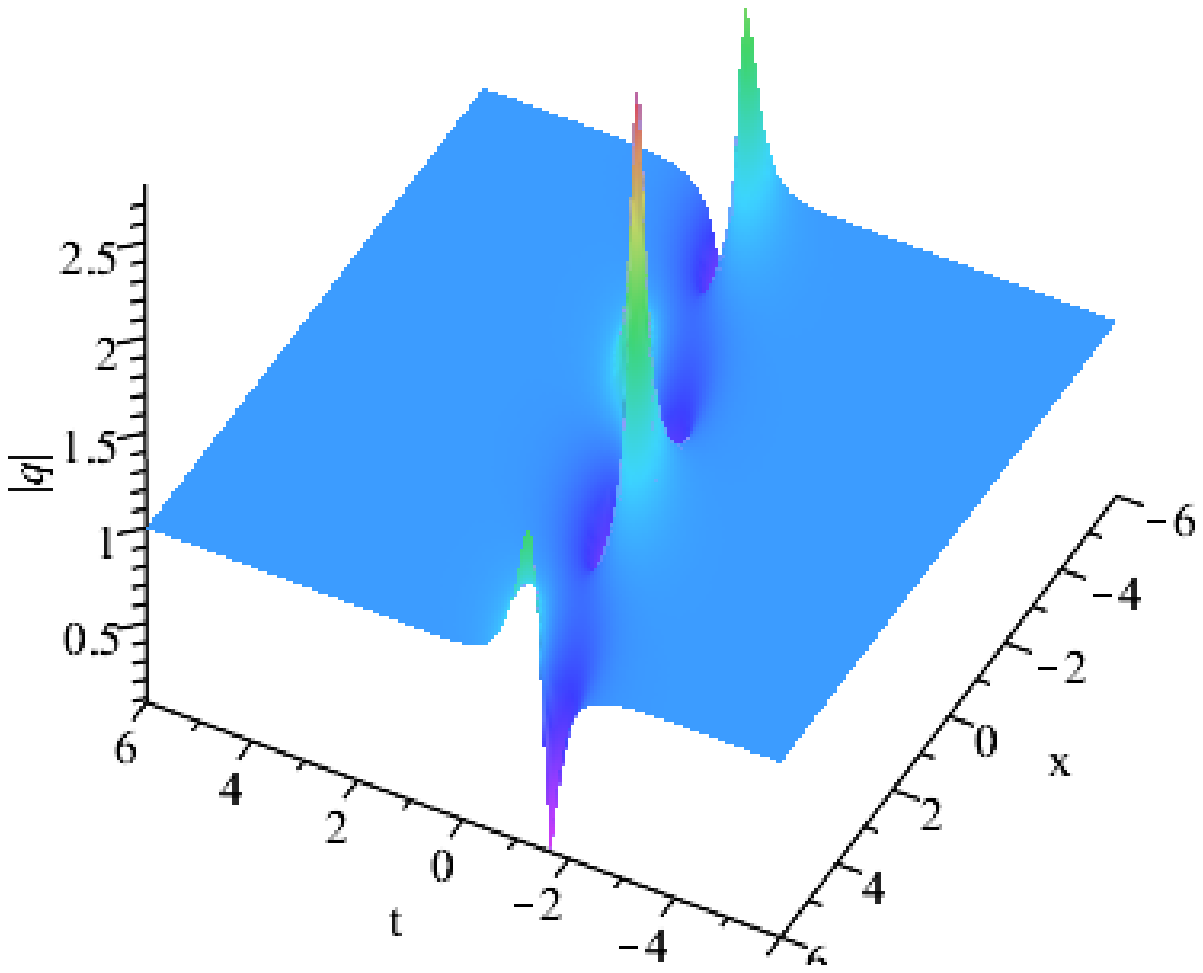}}}

 $\qquad\qquad(\textbf{a})\qquad \ \qquad\qquad\qquad\qquad\qquad \qquad\qquad\qquad(\textbf{b})$\\
\centerline{\noindent {\small \textbf{Figure 4.} Propagation of the solution Eq. \eqref{Jie-Q} with the the parameters (a) $z_{1}=\sqrt{2}e^{\frac{\pi i}{4}}$, (b) $z_{1}=\sqrt{2}e^{\frac{\pi i}{6}}$.}}

{\rotatebox{0}{\includegraphics[width=3.75cm,height=3.5cm,angle=0]{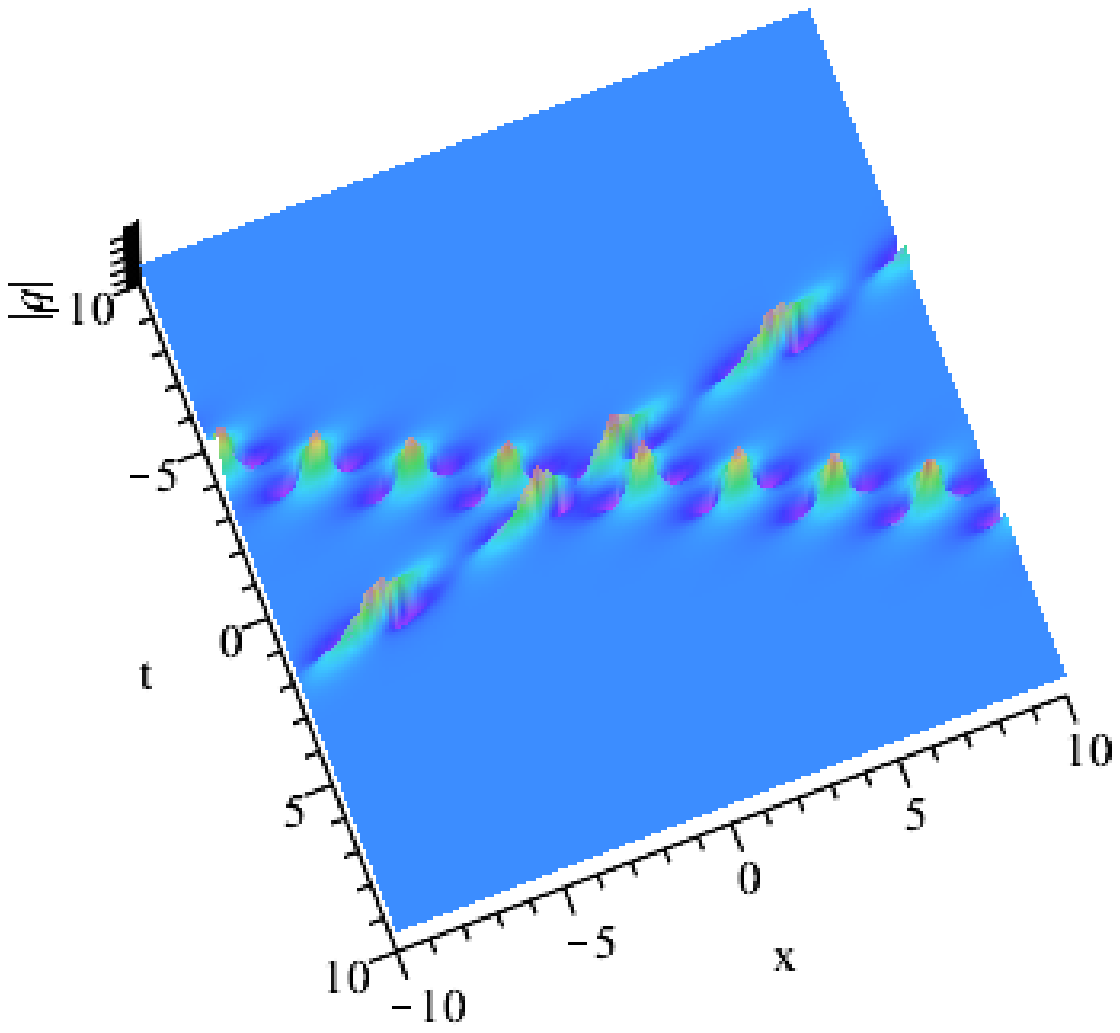}}}
\qquad\qquad\qquad\qquad
{\rotatebox{0}{\includegraphics[width=3.75cm,height=3.5cm,angle=0]{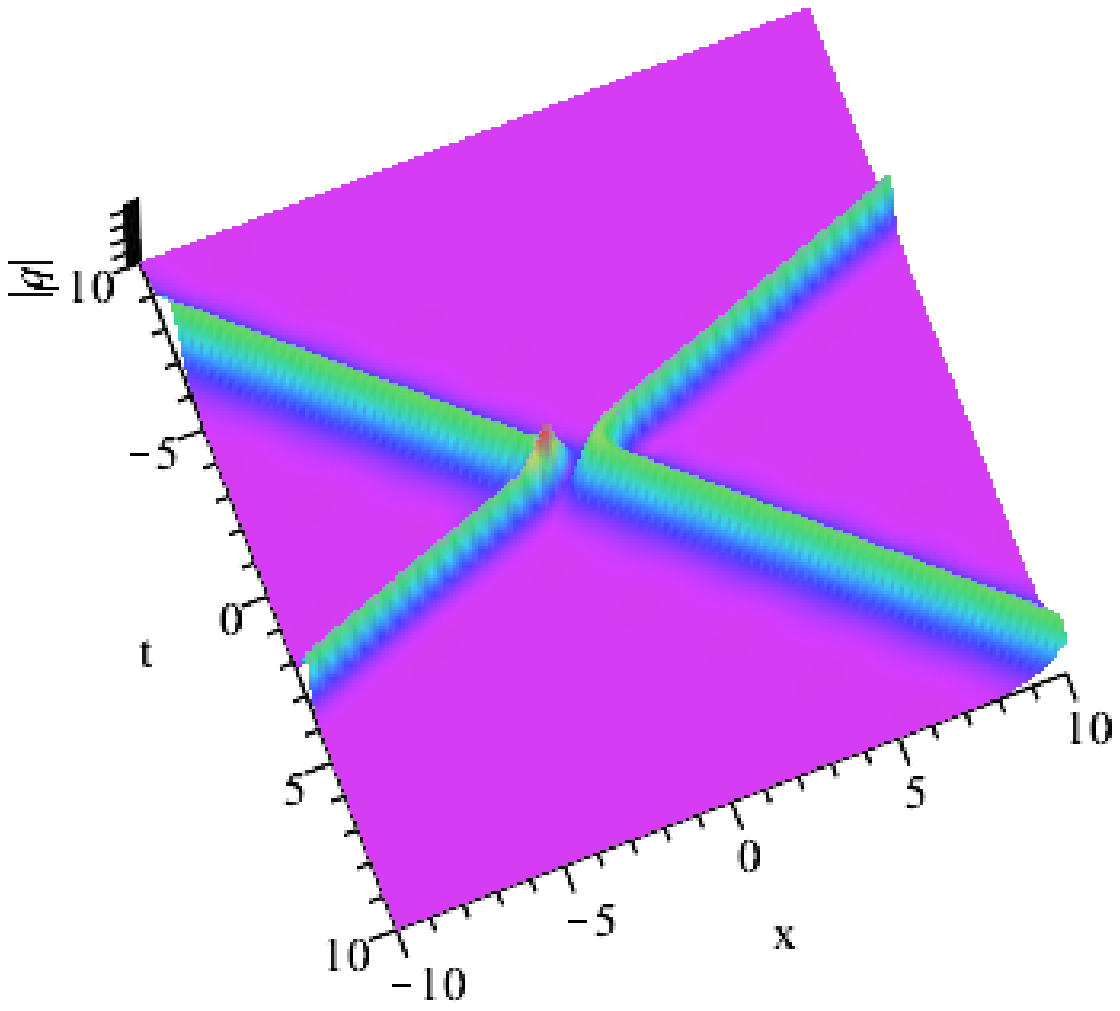}}}

 $\qquad\qquad(\textbf{a})\qquad \ \qquad\qquad\qquad\qquad\qquad \qquad\qquad\qquad(\textbf{b})$\\
\centerline{\noindent {\small \textbf{Figure 5.} Interaction of the solution Eq. \eqref{Jie-Q}.}}

\section{The ifoNLS equation with NZBCs and double zeros}
The discrete spectrum $\mathbb{Z}$ is the simple zero of scattering coefficient $s_{11}(z)$ and $s_{22}(z)$, which has already been discussed in previous work, then we will further discuss the case of double zero. In fact, the treatment of the direct scattering process of double zero is the same as that of simple zero, except for the treatment of discrete spectral points. Now assuming that the discrete spectrum $\mathbb{Z}$ is the double zeros of $s_{11}(z)$, i.e., $s_{11}(z_{n})=s'_{11}(z_{n})=0$ and $s''_{11}(z_{n})\neq0$. Although it is similar to the simple zero, there are many differences due to the existence of double zero, including trace formula, residue calculation, etc.

To obtain the residue conditions needed in inverse problem, we suppose that the notions
\begin{align}
\mathop{Res}_{z=z_{n}}[f(z)]=F_{-1},\quad \mathop{P_{-2}}_{z=z_{n}}[f(z)]=F_{-2},
\end{align}
denote the residue condition and the coefficient of $(z-z_{n})^{2}$ respectively in the Laurent expansion of a function $F(z)$ around $z_{n}$
\begin{align}
F(z)=\sum_{n=-\infty}^{\infty}\frac{F_{n}}{(z-z_{j})^{n}},\quad 0<z-z_{j}<R,
\end{align}
where $R$ stands for the associated radius of convergence.
According to the complex analysis, we know that if the functions $f(z)$ and $g(z)$ are analytic in the region $\Omega$, and the function  $g(z)$ is of a double zero $z_{n}\in\Omega$ but $f(z_{n})\neq0$. Then  the coefficients of the principal part of the Laurent expansion of $f(z)/g(z)$ can be expressed as
\begin{align}\label{T1}
\mathop{Res}_{z=z_{n}}\left[\frac{f(z)}{g(z)}\right]=\frac{2f'(z_{n})}{g''(z_{n})}-
\frac{2f(z_{n})g'''(z_{n})}{3(g''(z_{n}))^{2}},\quad
\mathop{P_{-2}}_{z=z_{n}}\left[\frac{f(z)}{g(z)}\right]=\frac{2f(z_{n})}{g''(z_{n})}.
\end{align}
Note that when the set $\mathbb{Z}$ is double zeros of $s_{11}(z)$ and $s_{22}(z)$, the Eq. \eqref{Q28} still holds, for simplicity, we still use the previous notation, namely the constant parameters $b_{-}(z_{n})$ and $b_{+}(z_{n}^{*})$.

To obtain residue conditions, we now investigate the linear relationship of Jost functions. From Eq. \eqref{s11}, one has
\begin{align}\label{T2}
[s_{11}(z)/\gamma]=Wr[\phi'_{+,1}(z),\phi_{-,2}(z)]+
Wr[\phi_{+,1}(z),\phi'_{-,2}(z)],
\end{align}
where the notion $'$ denotes the derivative of variable $z$. In terms of $z_{n}$ is the double zero of $s_{11}(z)$, thus
\begin{align}\label{T3}
Wr[\phi'_{+,1}(z_{n})-b_{-}(z_{n})\phi'_{-,2}(z_{n}),\phi_{-,2}(z_{n})]=0.
\end{align}
Obviously, there exits a constant $d_{-}(z_{n})$ such that
\begin{align}\label{T4}
\phi'_{+,1}(x,t;z_{n})&=d_{-}(z_{n})\phi_{-,2}(x,t;z_{n})+
b_{-}(z_{n})\phi'_{-,2}(x,t;z_{n}).
\end{align}
Resorting  to the expression
\begin{align}\label{T5}
\phi_{+,2}(x,t;z_{n}^{*})&=b_{+}(z_{n}^{*})\phi_{-,1}(x,t;z_{n}^{*}),
\end{align}
the another linear relationship can be  similarly derived by
\begin{align}\label{T6}
\phi'_{+,2}(x,t;z_{n}^{*})&=d_{+}(z_{n}^{*})\phi_{-,1}(x,t;z_{n}^{*})+
b_{+}(z_{n}^{*})\phi'_{-,1}(x,t;z_{n}^{*}),
\end{align}
where $d_{+}(z_{n}^{*})$ is a constant parameter.
In terms of the discrete spectrum $z_{n}$ and $z_{n}^{*}$ are the zeros of $s_{11}(z)$ and $s_{22}(z)$, respectively, thus one has the residue conditions
\addtocounter{equation}{1}
\begin{align}\label{Res1}
\mathop{P_{-2}}_{z=z_{n}}\left[\frac{\phi_{+,1}(x,t;z)}{s_{11}(z)}\right]&=
\frac{2\phi_{+,1}(z_{n})}{s''_{11}(z_{n})}=\frac{2b_{-}(z_{n})}{s''_{11}(z_{n})}
\phi_{-,2}(z_{n})=A_{-}[z_{n}]\phi_{-,2}(z_{n}), \tag{\theequation a}\\
\mathop{Res}_{z=z_{n}}\left[\frac{\phi_{+,1}(x,t;z)}{s_{11}(z)}\right]
&=\frac{2b_{-}(z_{n})}{s''_{11}(z_{n})}\left[\phi'_{-,2}(z_{n})+
\left(\frac{d_{-}(z_{n})}{b_{-}(z_{n})}-\frac{s'''_{11}(z_{n})}{3s''_{11}(z_{n})}
\right)\phi_{-,2}(z_{n})\right],\tag{\theequation b} \label{t2}\\
\mathop{P_{-2}}_{z=z_{n}^{*}}\left[\frac{\phi_{+,2}(x,t;z)}{s_{22}(z)}\right]&=
\frac{2\phi_{+,2}(z_{n}^{*})}{s''_{22}(z_{n}^{*})}=
\frac{2b_{+}(z_{n}^{*})}{s''_{22}(z_{n}^{*})}\phi_{-,1}(z_{z_{n}}^{*})
=A_{+}[z_{n}^{*}]\phi_{-,1}(z_{n}^{*}), \tag{\theequation c}\\
\mathop{Res}_{z=z_{n}^{*}}\left[\frac{\phi_{+,2}(x,t;z)}{s_{22}(z)}\right]
&=\frac{2b_{+}(z_{n}^{*})}{s''_{22}(z_{n}^{*})}\left[\phi'_{-,1}(z_{n}^{*})+
\left(\frac{d_{+}(z_{n}^{*})}{b_{+}(z_{n}^{*})}-
\frac{s'''_{22}(z_{n}^{*})}{3s''_{22}(z_{n}^{*})}
\right)\phi_{-,1}(z_{n}^{*})\right],\tag{\theequation d} \label{t4}
\end{align}
with the notions
\begin{align}\label{T7}
A_{-}[z_{n}]=\frac{2b_{-}(z_{n})}{s''_{11}(z_{n})},\quad
B_{-}[z_{n}]=\frac{d_{-}(z_{n})}{b_{-}(z_{n})}-
\frac{s'''_{11}(z_{n})}{3s''_{11}(z_{n})},\\
A_{+}[z_{n}^{*}]=\frac{2b_{+}(z_{n}^{*})}{s''_{22}(z_{n}^{*})},\quad
B_{+}[z_{n}^{*}]=\frac{d_{+}(z_{n}^{*})}{b_{+}(z_{n}^{*})}-
\frac{s'''_{22}(z_{n}^{*})}{3s''_{22}(z_{n}^{*})}.
\end{align}
It follows based on the asymmetries of Jost functions and scattering matrix as well as Eqs. \eqref{T4} and \eqref{T6} that
\begin{align}
\begin{split}
A_{-}[z_{n}]&=-A_{+}^{*}[z_{n}^{*}]=\frac{z_{n}^{4}q_{-}^{*}}
{q_{0}^{4}q_{-}}A_{+}\left(-\frac{q_{0}^{2}}{z_{n}}\right)=-\frac{z_{n}^{4}q_{-}^{*}}
{q_{0}^{4}q_{-}}A_{-}\left(-\frac{q_{0}^{2}}{z_{n}^{*}}\right),
\quad\quad z_{n}\in \mathbb{Z}\cap D^{-},\\
B_{-}[z_{n}]&=B_{+}^{*}[z_{n}^{*}]=\frac{q_{0}^{2}}
{z_{n}^{2}}B_{+}\left(-\frac{q_{0}^{2}}{z_{n}}\right)+\frac{2}{z_{n}}=\frac{q_{0}^{2}}
{z_{n}^{2}}B_{-}^{*}\left(-\frac{q_{0}^{2}}{z_{n}^{*}}\right)+\frac{2}{z_{n}}.
\quad z_{n}\in \mathbb{Z}\cap D^{-}.
\end{split}
\end{align}
Similarly, for the discrete spectrum $-\frac{q_{0}^{2}}{z_{n}}$ and $\frac{-q_{0}^{2}}{z_{n}^{*}}$, one can get these residue conditions
\begin{align*}
\mathop{P_{-2}}_{z=\frac{-q_{0}^{2}}{z_{n}^{*}}}\left[\frac{\phi_{+,1}(z)}
{s_{11}(z)}\right]&=\frac{2\phi_{+,1}\left(\frac{-q_{0}^{2}}{z_{n}^{*}}\right)}
{s''_{11}\left(\frac{-q_{0}^{2}}{z_{n}^{*}}\right)}=
\frac{2b_{-}\left(\frac{-q_{0}^{2}}{z_{n}^{*}}\right)}
{s''_{11}\left(\frac{-q_{0}^{2}}{z_{n}^{*}}\right)}
\phi_{-,2}\left(\frac{-q_{0}^{2}}{z_{n}^{*}}\right)=
A_{-}\left[\frac{-q_{0}^{2}}{z_{n}^{*}}\right]
\phi_{-,2}\left(\frac{-q_{0}^{2}}{z_{n}^{*}}\right), \\
\mathop{Res}_{z=\frac{-q_{0}^{2}}{z_{n}^{*}}}\left[\frac{\phi_{+,1}(z)}{s_{11}(z)}\right]
&=\frac{2b_{-}\left(\frac{-q_{0}^{2}}{z_{n}^{*}}\right)}
{s''_{11}\left(\frac{-q_{0}^{2}}{z_{n}^{*}}\right)}
\left[\phi'_{-,2}\left(\frac{-q_{0}^{2}}{z_{n}^{*}}\right)+
\left(\frac{d_{-}\left(\frac{-q_{0}^{2}}{z_{n}^{*}}\right)}
{b_{-}\left(\frac{-q_{0}^{2}}{z_{n}^{*}}\right)}-
\frac{s'''_{11}\left(\frac{-q_{0}^{2}}{z_{n}^{*}}\right)}
{3s''_{11}\left(\frac{-q_{0}^{2}}{z_{n}^{*}}\right)}\right)
\phi_{-,2}\left(\frac{-q_{0}^{2}}{z_{n}^{*}}\right)\right], \\
\mathop{P_{-2}}_{z=-\frac{q_{0}^{2}}{z_{n}}}\left[\frac{\phi_{+,2}(z)}{s_{22}(z)}\right]&=
\frac{2\phi_{+,2}\left(-\frac{q_{0}^{2}}{z_{n}}\right)}
{s''_{22}\left(-\frac{q_{0}^{2}}{z_{n}}\right)}=
\frac{2b_{+}\left(-\frac{q_{0}^{2}}{z_{n}}\right)}
{s''_{22}\left(-\frac{q_{0}^{2}}{z_{n}}\right)}
\phi_{-,1}\left(-\frac{q_{0}^{2}}{z_{n}}\right)
=A_{+}\left[-\frac{q_{0}^{2}}{z_{n}}\right]
\phi_{-,1}\left(-\frac{q_{0}^{2}}{z_{n}}\right), \\
\mathop{Res}_{z=-\frac{q_{0}^{2}}{z_{n}}}\left[\frac{\phi_{+,2}(z)}{s_{22}(z)}\right]
&=\frac{2b_{+}\left(-\frac{q_{0}^{2}}{z_{n}}\right)}
{s''_{22}\left(-\frac{q_{0}^{2}}{z_{n}}\right)}
\left[\phi'_{-,1}\left(-\frac{q_{0}^{2}}{z_{n}}\right)+
\left(\frac{d_{+}\left(-\frac{q_{0}^{2}}{z_{n}}\right)}
{b_{+}\left(-\frac{q_{0}^{2}}{z_{n}}\right)}-
\frac{s'''_{22}\left(-\frac{q_{0}^{2}}{z_{n}}\right)}
{3s''_{22}\left(-\frac{q_{0}^{2}}{z_{n}}\right)}
\right)\phi_{-,1}\left(-\frac{q_{0}^{2}}{z_{n}}\right)\right].
\end{align*}
Based on the discrete spectrum $\mathbb{Z}=\left\{\eta_{n}, \hat{\eta}_{n}\right\}$ and Eqs. \eqref{Q16} and \eqref{Matr}, the above residue conditions can be  taken the more compact form
\addtocounter{equation}{1}
\begin{align}\label{Residue1}
\mathop{P_{-2}}_{z=\eta_{n}}M_{1}^{-}(z)=\mathop{P_{-2}}_{z=\eta_{n}}
\left[\frac{u_{+,1}(z)}{s_{11}(z)}\right]&=\frac{2u_{+,1}(\eta_{n})}
{s''_{11}(\eta_{n})}=A_{-}[\eta_{n}]e^
{2i\theta(\eta_{n})}u_{-,2}(\eta_{n}),\tag{\theequation a}\\
\mathop{P_{-2}}_{z=\hat{\eta}_{n}}M_{2}^{+}(z)=\mathop{P_{-2}}_{z=\hat{\eta}_{n}}
\left[\frac{u_{+,2}(z)}{s_{22}(z)}\right]&=\frac{2u_{+,2}(\hat{\eta}_{n})}
{s''_{22}(\hat{\eta}_{n})}=A_{+}[\hat{\eta}_{n}]e^
{-2i\theta(\hat{\eta}_{n})}u_{-,1}(\hat{\eta}_{n}),\tag{\theequation b}\\
\mathop{Res}_{z=\eta_{n}}M_{1}^{-}(z)=\mathop{Res}_{z=\eta_{n}}
\left[\frac{u_{+,1}(z)}{s_{11}(z)}\right]&=A_{-}[\eta_{n}]
e^{2i\theta(\eta_{n})}\left[u'_{-,2}(\eta_{n})+
\Delta_{n}^{-}u_{-,2}(\eta_{n}))\right],\tag{\theequation c}\\
\mathop{Res}_{z=\hat{\eta}_{n}}M_{2}^{+}(z)=\mathop{Res}_{z=\hat{\eta}_{n}}
\left[\frac{u_{+,2}(z)}{s_{22}(z)}\right]&=A_{+}[\hat{\eta}_{n}]
e^{-2i\theta(\hat{\eta}_{n})}\left[u'_{-,1}
(\hat{\eta}_{n})+\Delta_{n}^{+}
u_{-,1}(\hat{\eta}_{n}))\right].
\tag{\theequation d} \label{Residue4}
\end{align}
with
\begin{align}\label{4.14}
\begin{split}
A_{+}[\hat{\eta}_{n}]=\frac{2b_{+}(\hat{\eta}_{n})}{s''_{22}(\hat{\eta}_{n})},\quad
\Delta_{n}^{+}=B_{+}[\hat{\eta}_{n}]-2i\theta'(\hat{\eta}_{n}),\\
A_{-}[\eta_{n}]=\frac{2b_{-}(\eta_{n})}{s''_{11}(\eta_{n})},\quad
\Delta_{n}^{-}=B_{-}[\eta_{n}]+2i\theta'(\eta_{n}).
\end{split}
\end{align}
\subsection{The exact formula of the ifoNLS with double zeros}
The RH problem Eq. \eqref{Matr} constructed in previous work with single zeros is still applicable to the case of double zeros, but in order to solve the RH problem, it is necessary to subtract the residue conditions generated by the corresponding zeros to ensure the analyticity. By using the properties of projection operators Eq. \eqref{Cauchy}, namely when the functions $f_{\pm}(k)$ are analytic in upper and lower of $k$-plane and as $|k|\rightarrow\infty$ the functions $f_{\pm}(k)\rightarrow 0$, one can get
\begin{align}\label{T8}
P^{\pm}[f_{\pm}(k)]=\pm f_{\pm}(k),\quad P^{\pm}[f_{\mp}(k)]=0.
\end{align}
Combining with the asymmetry and the analyticity of the function $M(x,t;z)$, we can derive via Eq. \eqref{T8} and Plemelj's formulae
\begin{align}\label{T9}
\begin{split}
M(x,t;z)=&\mathbb{I}-\frac{i}{z}\sigma_{3}Q_{-}+\sum_{n=1}^{2N}\left\{
\frac{\mathop{Res}\limits_{z=\hat{\eta}_{n}}M^{+}}{z-\hat{\eta}_{n}}
+\frac{\mathop{P_{-2}}\limits_{z=\hat{\eta}_{n}}M^{+}}{(z-\hat{\eta}_{n})^{2}}
+\frac{\mathop{Res}\limits_{z=\eta_{n}}M^{-}}{z-\eta_{n}}
+\frac{\mathop{P_{-2}}\limits_{z=\eta_{n}}M^{-}}{(z-\eta_{n})^{2}}\right\} \\
&+\frac{1}{2\pi i}\int_{\Sigma}\frac{M^{-}(x,t;\xi)G(x,t;\xi)}{\xi-z}\,d\xi,
\quad z\in\mathbb{C}\setminus\Sigma.
\end{split}
\end{align}
To give the exact expression of $M(x,t;z)$, we need to calculate the residue conditions of Eq. \eqref{T9}. Resorting to Eq. \eqref{Matr}, one has
\begin{align}\label{T10}
\begin{split}
\mathop{Res}_{z=\eta_{n}}[M^{-}]=\left(\mathop{Res}_{z=\eta_{n}}
\left[\frac{\mu_{+,1}(x,t;z)}{s_{11}(z)}\right],0\right), \quad \mathop{P_{-2}}_{z=\eta_{n}}M^{-}=\left(\mathop{P_{-2}}_{z=\eta_{n}}
\left[\frac{\mu_{+,1}(x,t;z)}{s_{11}(z)}\right],0\right),  \\
\mathop{Res}_{z=\hat{\eta}_{n}}[M^{+}]=\left(0,\mathop{Res}_{z=\hat{\eta}_{n}}
\left[\frac{\mu_{+,2}(x,t;z)}{s_{22}(z)}\right]\right), \quad \mathop{P_{-2}}_{z=\hat{\eta}_{n}}M^{+}=\left(0,\mathop{P_{-2}}_{z=\hat{\eta}_{n}}
\left[\frac{\mu_{+,2}(x,t;z)}{s_{22}(z)}\right]\right),
\end{split}
\end{align}
further
\begin{align}\label{T11}
\begin{split}
&\frac{\mathop{Res}\limits_{z=\hat{\eta}_{n}}M^{+}}{z-\hat{\eta}_{n}}
+\frac{\mathop{P_{-2}}\limits_{z=\hat{\eta}_{n}}M^{+}}{(z-\hat{\eta}_{n})^{2}}
+\frac{\mathop{Res}\limits_{z=\eta_{n}}M^{-}}{z-\eta_{n}}
+\frac{\mathop{P_{-2}}\limits_{z=\eta_{n}}M^{-}}{(z-\eta_{n})^{2}}=\\&
\left(\nabla_{n}^{-}(z)\left[u'_{-,2}(\eta_{n})+
\left(\Delta_{n}^{-}+\frac{1}{z-\eta_{n}}\right)u_{-,2}(\eta_{n})\right],
\nabla_{n}^{+}(z)\left[u'_{-,1}(\hat{\eta}_{n})+\left(\Delta_{n}^{+}+
\frac{1}{z-\hat{\eta}_{n}}\right)u_{-,1}(\hat{\eta}_{n})\right]\right),
\end{split}
\end{align}
with
\begin{align}\label{4.19}
\nabla_{n}^{-}(z)=\frac{A_{-}[\eta_{n}]}{z-\eta_{n}}e^{2i\theta(\eta_{n})},
\nabla_{n}^{+}(z)=\frac{A_{+}[\hat{\eta}_{n}]}{z-\hat{\eta}_{n}}
e^{-2i\theta(\hat{\eta}_{n})}.
\end{align}
We next evaluate the second column of the function $M(x,t;z)$ at $z=\eta_{s}$, $s=1,2,\cdots,2N$
\begin{align}\label{T12}
\begin{split}
u_{-,2}(x,t;\eta_{s})=&\left(\begin{array}{cc}
                       -\frac{iq_{-}}{\eta_{s}} \\
                        1
                     \end{array}\right)
+\frac{1}{2\pi i}\int_{\Sigma}\frac{(M^{-}G)_{2}(\xi)}{\xi-\eta_{\xi}}\,d\xi \\ &+\sum_{n=1}^{2N}H_{n}^{+}(\eta_{s})
\left[u'_{-,1}(x,t;\hat{\eta}_{n})+\left(\Delta_{n}^{+}+
\frac{1}{\eta_{s}-\hat{\eta}_{n}}\right)
u_{-,1}(x,t;\hat{\eta}_{n})\right].
\end{split}
\end{align}
Furthermore
\begin{align}\label{T13}
\begin{split}
u'_{-,2}(x,t;\eta_{s})=&\left(\begin{array}{cc}
                       \frac{iq_{-}}{\eta_{s}^{2}} \\
                        0
                     \end{array}\right)
+\frac{1}{2\pi i}\int_{\Sigma}\frac{(M^{-}G)_{2}(\xi)}{(\xi-\eta_{s})^{2}}\,
d\xi \\
&-\sum_{n=1}^{2N}\frac{\nabla_{n}^{+}(\eta_{s})}{\eta_{s}-\hat{\eta}_{n}}
\left[u'_{-,1}(x,t;\hat{\eta}_{n})+\left(\Delta_{n}^{+}+
\frac{2}{\eta_{s}-\hat{\eta}_{n}}\right)
u_{-,1}(x,t;\hat{\eta}_{n})\right].
\end{split}
\end{align}
It follows from Eq. \eqref{Q23} and the set $\mathbb{Z}$ that
\begin{align}\label{T14}
\begin{split}
u_{-,2}(\eta_{s})&=-\frac{iq_{-}}{\eta_{s}}u_{-,1}(\hat{\eta}_{s}),\\
u'_{-,2}(\eta_{s})&=\frac{iq_{-}}{\eta_{s}^{2}}u_{-,1}(\hat{\eta}_{s})-
\frac{iq_{0}^{2}q_{-}}{\eta_{s}^{3}}u'_{-,1}(\hat{\eta}_{s}).
\end{split}
\end{align}
Inserting Eq. \eqref{T14} into Eqs. \eqref{T12} and \eqref{T13}, one has
\begin{align}\label{T15}
\begin{split}
&\sum_{n=1}^{2N}\left(\nabla^{+}_{n}(\eta_{s})u'_{-,1}(x,t;\hat{\eta}_{n})+
\left[\nabla^{+}_{n}(\eta_{s})\left(\Delta^{+}_{n}+\frac{1}{\eta_{s}-\hat{\eta}_{n}}\right)+
\frac{iq_{-}}{\eta_{s}}\delta_{sn}\right]u_{-,1}(\hat{\eta}_{n})\right)\\
&+\left(\begin{array}{cc}
        -\frac{iq_{-}}{\eta_{s}} \\
         1
       \end{array}\right)
+\frac{1}{2\pi i}\int_{\Sigma}\frac{(M^{-}G)_{2}(\zeta)}{\zeta-\eta_{s}}\,d\zeta=0,\\
&\sum_{n=1}^{2N}\frac{\nabla_{n}^{+}(\eta_{k})}{\eta_{s}-\hat{\eta}_{n}}\left(\left[
\left(\Delta_{n}^{+}+\frac{2}{\eta_{s}-\hat{\eta}_{n}}\right)
+\frac{iq_{-}}{\eta_{k}^{2}}\delta_{sn}\right]u_{-,1}(\hat{\eta}_{n})-
\frac{iq_{-}q_{0}^{2}}{\eta_{s}^{3}}\delta_{sn}u'_{-,1}(\hat{\eta}_{n})\right) \\
&-\left(\begin{array}{cc}
        \frac{iq_{-}}{\eta_{s}^{2}} \\
         0
       \end{array}\right)
-\frac{1}{2\pi i}\int_{\Sigma}\frac{(M^{-}G)_{2}(\xi)}{(\xi-\eta_{s})^{2}}\,d\xi=0,
\end{split}
\end{align}
which generates an algebraic system including the elements needed in constructing solutions. Obviously the expressions of $u_{-,1}(x,t;\hat{\eta}_{n})$ and $u'_{-,1}(x,t;\hat{\eta}_{n})$ can be derived by Eq. \eqref{T15}, then inserting the obtained results into Eq. \eqref{T14}, one has the formula of $u_{-,2}(x,t;\eta_{n})$ and $u'_{-,2}(x,t;\eta_{n})$. Finally the expression of $M(x,t;z)$ can be given by substituting these expressions into Eq. \eqref{T9}. Therefore the following  proposition can be obtained.
\begin{prop}
The solution to the ifoNLS equation  with reflection-less potential  is written as under double zeros condition
\begin{align}\label{T16}
\begin{split}
q(x,t)=-q_{-}-&i\sum_{n=1}^{2N}A_{+}[\hat{\eta}_{n}]e^{-2i\theta(x,t;\hat{\eta}_{n})}
[u'_{-,1,1}(x,t;\hat{\eta}_{n})+\Delta_{n}^{+}u_{-,1,1}(x,t;\hat{\eta}_{n})],
\end{split}
\end{align}
with  the functions $u'_{-,1,1}(\hat{\eta}_{n})$ and $u_{-,1,1}(\hat{\eta}_{n})$ are determined by
\begin{align}\label{T17}
\begin{split}
&\sum_{n=1}^{2N}\left(\nabla^{+}_{n}(\eta_{s})u'_{-,1,1}(\hat{\eta}_{n})+
\left[\nabla^{+}_{n}(\eta_{s})\left(\Delta^{+}_{n}+\frac{1}{\eta_{s}-
\hat{\eta}_{n}}\right)+
\frac{iq_{-}}{\eta_{s}}\delta_{sn}\right]u_{-,1,1}(\hat{\eta}_{n})\right)
=\frac{iq_{-}}{\eta_{s}},\\
&\sum_{n=1}^{2N}\frac{\nabla_{n}^{+}(\eta_{k})}{\eta_{s}-\hat{\eta}_{n}}\left(\left[
\left(\Delta_{n}^{+}+\frac{2}{\eta_{s}-\hat{\eta}_{n}}\right)
+\frac{iq_{-}}{\eta_{k}^{2}}\delta_{sn}\right]u_{-,1,1}(\hat{\eta}_{n})-
\frac{iq_{-}q_{0}^{2}}{\eta_{s}^{3}}\delta_{sn}u'_{-,1,1}(\hat{\eta}_{n})\right)
=\frac{iq_{-}}{\eta_{s}^{2}},
\end{split}
\end{align}
which is equivalent to Theorem 1.2,
note that
\begin{align}
\begin{split}
\tilde{G}&=\left(\begin{array}{cc}
                  \tilde{G}^{11} & \tilde{G}^{12} \\
                  \tilde{G}^{21} & \tilde{G}^{22} \\
                \end{array}
              \right),\quad
\tilde{G}^{(sj)}=(\tilde{g}_{kn}^{(sj)})_{(2N)\times(2N)},\\
 \tilde{g}_{kn}^{(11)}&=\nabla_{n}^{+}(\eta_{k})\left(\Delta_{n}^{+}+\frac{1}{\eta_{k}-
 \hat{\eta}_{n}}\right)+\frac{iq_{-}}{\eta_{k}}\delta_{k,n}, \\ \tilde{g}_{kn}^{(12)}&=\nabla_{n}^{+}(\eta_{k}),\quad
 \tilde{g}_{kn}^{(21)}=\frac{\nabla_{n}^{+}(\eta_{k})}{\eta_{k}-
 \hat{\eta}_{n}}\left(\Delta_{n}^{+}
 +\frac{2}{\eta_{k}-\hat{\eta}_{n}}\right)+\frac{iq_{-}}
 {\eta_{k}^{2}}\delta_{k,n},\\
 \tilde{g}_{kn}^{(22)}&=\frac{\nabla_{n}^{+}(\eta_{k})}{\eta_{k}-\hat{\eta}_{n}}-
 \frac{iq_{-}q_{0}^{2}}{\eta_{k}^{3}}\delta_{k,n},\\
\omega_{n}^{(1)}&=A_{+}[\hat{\eta_{n}}]e^{-2i\theta(\hat{\eta_{n}})}\Delta_{n}^{+},\quad
\omega_{n}^{(2)}=A_{+}[\hat{\eta_{n}}]e^{-2i\theta(\hat{\eta_{n}})},\\
\nu_{n}^{(1)}&=\frac{iq_{-}}{\eta_{n}}, \quad \nu_{n}^{(2)}=\frac{iq_{-}}{\eta_{n}^{2}},\quad
y=(y_{1}^{(1)}, \cdots, y_{2N}^{(1)}, y_{1}^{(2)}, \cdots, y_{2N}^{(2)})^{T},\\
y_{n}^{(1)}&=u_{-,1,1}(\hat{\eta}_{n}),\quad  y_{n}^{(2)}=u'_{-,1,1}(\hat{\eta}_{n}).
\end{split}
\end{align}
\end{prop}
In order to further observe the dynamic behavior of the solution Eq. \eqref{T20}, we choose appropriate parameters to give the following solution image.

{\rotatebox{0}{\includegraphics[width=3.75cm,height=3.5cm,angle=0]{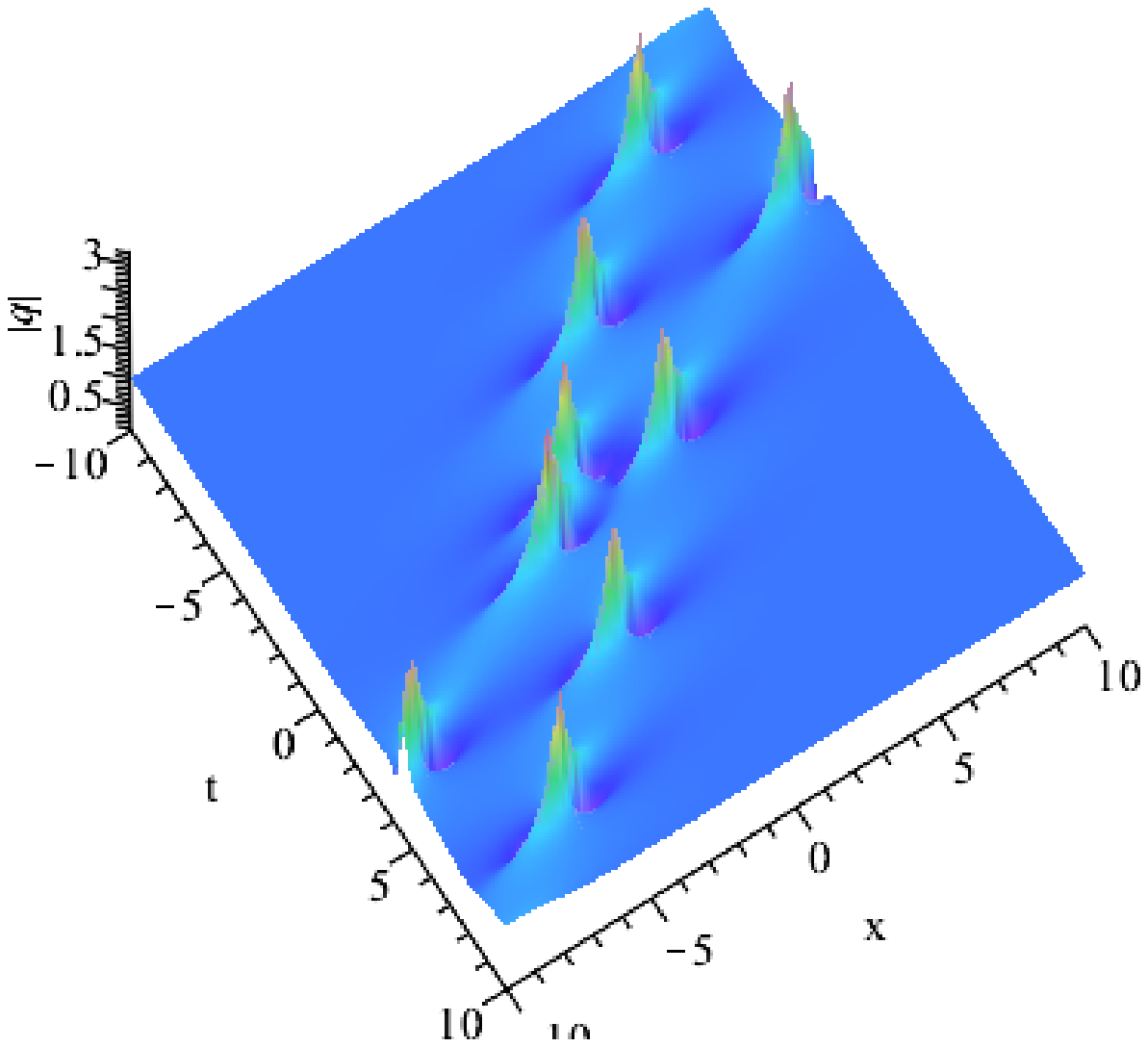}}}
\qquad\qquad\qquad\qquad
{\rotatebox{0}{\includegraphics[width=3.75cm,height=3.5cm,angle=0]{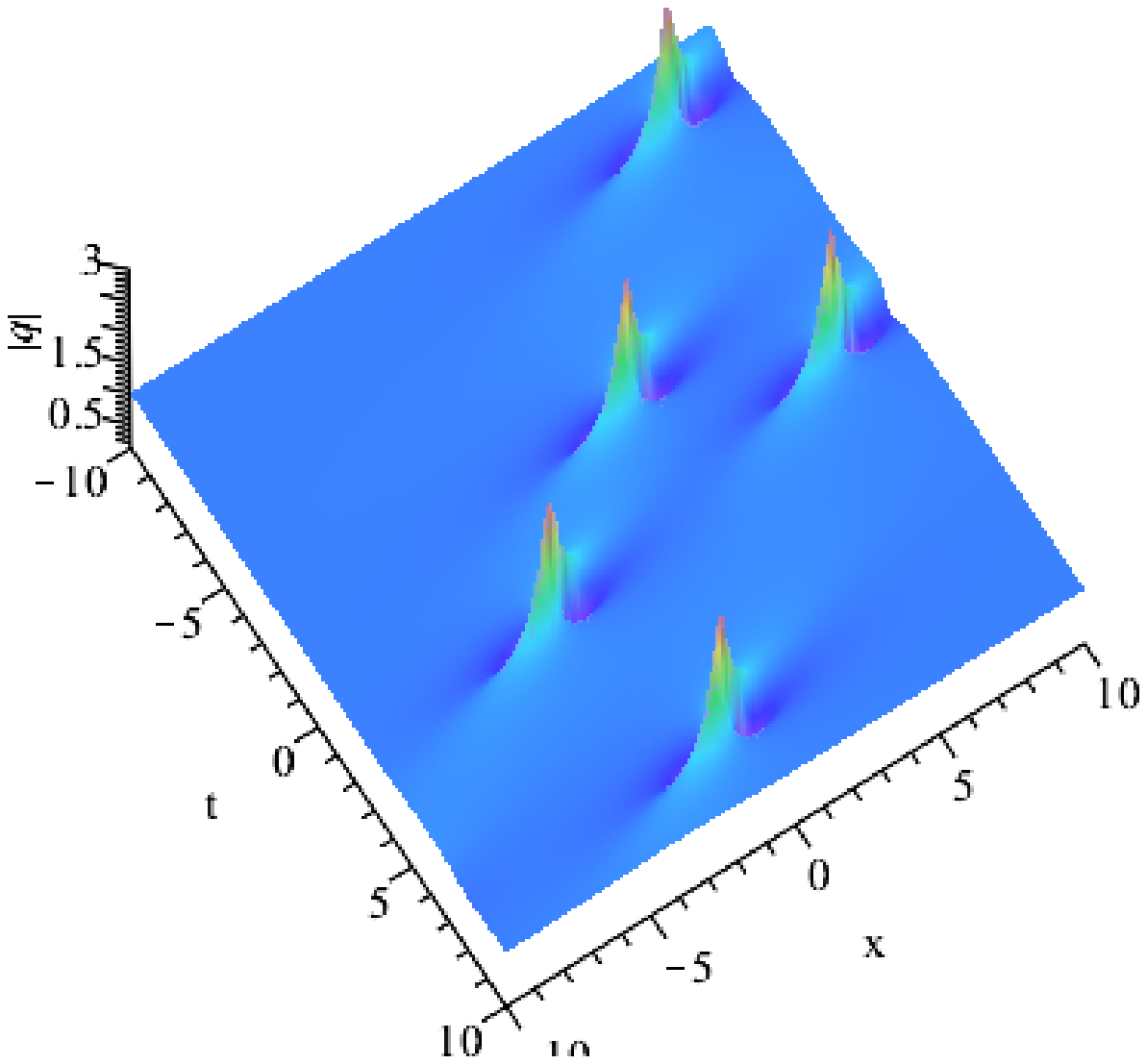}}}

 $\qquad\qquad(\textbf{a})\qquad \ \qquad\qquad\qquad\qquad\qquad \qquad\qquad\qquad(\textbf{b})$\\
\centerline{\noindent {\small \textbf{Figure 6.} The propagation of the solution Eq. \eqref{T20}.} $q_{0}=1$, $\epsilon=0.01$, ($a$): $z_{1}=1.5i$, ($b$): $z_{1}=1.25i$.}

\section{Conclusions and discussions}
In this work, we systematically study the inhomogeneous fifth-order nonlinear Schr\"{o}dinger equation with nonzero boundary condition. Based on the theory of inverse scattering, the RH problem is constructed and the exact solutions of the equation Eq. \eqref{Q1}  are given under the condition of simple zeros and double zeros.
In the process of solving RH problem, in order to ensure that the two ends of the equation about function $M^{\pm}$ are analytical in the corresponding region, the main idea is to subtract the residue conditions  generated by simple poles and two poles and asymptotic behavior, and we finally give the formal solution of RH problem through Plemelj's formula.
 Therefore, we can further discuss higher order zeros (greater than the double zeros), such as $N$-order zeros, and give more general solutions. At present, there is no literature about discussing higher order zeros, which is also a problem worth thinking.

\section*{Appendix A: Proof of Theorem 2.1.}

We consider the first integral equations of  Eq. \eqref{Q17}
\begin{equation*}
u_{-}(x,t;z)=Y_{-}+\int_{-\infty}^{x}Y_{-}e^{-i\lambda(x-y)\sigma_{3}}Y_{-}^{-1}\Delta Q_{-}(y,t)u_{-}(y,t;z)e^{i\lambda(x-y)\sigma_{3}}\, dy. \eqno(A1)
\end{equation*}
Obviously from the limits of Eq.(A1), one knows that $x-y>0$. Note that
\begin{equation*}
e^{-i\theta\sigma_{3}}Qe^{i\theta\sigma_{3}}=\left(\begin{array}{cc}
q_{11} & e^{-2i\theta}q_{12}\\
e^{2i\theta}q_{21} & q_{22}\\
\end{array}\right). \eqno(A2)
\end{equation*}
Eq. (A1) can be written as
\begin{equation*}
Y_{-}^{-1}u_{-}(x,t;z)=I+\int_{-\infty}^{x}e^{-i\lambda(x-y)\sigma_{3}}Y_{-}^{-1}\Delta Q_{-}(y,t)u_{-}(y,t;z)e^{i\lambda(x-y)\sigma_{3}}\, dy. \eqno(A3)
\end{equation*}
The matrix Eq. (A3) is divided into columns, and suppose $W(x,z)=Y_{-}^{-1}u_{-}$, we have
\begin{equation*}
w(x,z)=\left(\begin{array}{c}
1\\
0\\
\end{array}\right)+\int_{-\infty}^{x}G(x-y,z)\Delta Q_{-}(y,t)Y_{-}(y,t;z)w(y,z)\, dy,\eqno(A4)
\end{equation*}
where
\begin{equation*}
G(\theta,z)=diag(1,e^{2i\lambda(z)\theta})Y_{-}^{-1}(z)=
\frac{1}{\gamma}\left(\begin{array}{cc}
1 & \frac{iq_{-}}{z}\\
\frac{iq_{-}^{*}}{z}e^{2i\lambda(z)\theta} & e^{2i\lambda(z)\theta}\\
\end{array}\right).\eqno(A5)
\end{equation*}
Then introducing a Neumann series representation for $w$,
\begin{align*}
w(x,z)&=\sum_{n=0}^{\infty}w^{n},\\
w^{0}=\left(\begin{array}{c}
1\\
0\\
\end{array}\right),\quad
w^{n+1}(x,z)&=\int_{-\infty}^{x}C(x,y,z)w^{n}(y,z)\, dy,
\end{align*}
 similarly \cite{NZBC-14}, we can prove that for all $\epsilon>0$, if $q(x)-q_{-}\in L^{1}(-\infty,a]$ for $a\in\mathbb{R}$, the Neumann series converges absolutely and uniformly with respect to $x\in(-\infty, a)$ and $z\in D^{-}_{\epsilon}$ where $D^{-}_{\epsilon}(z_{0})=\left\{z\in\mathbb{C}:|z-z_{0}|<\epsilon q_{0}\right\}$. Due to uniformly series convergent series of analytic functions converges to an analytic function, we know that the column of the Jost functions are analytic in the domain. We only take the first column of the matrix as an example to prove it in detail, and the rest are similar to verifiable cases.
\begin{align*}
u_{-}(x,t;z): \quad
e^{-i\lambda(x-y)\sigma_{3}}Me^{i\lambda(x-y)\sigma_{3}}=\left(\begin{array}{cc}
m_{11} & e^{-2i\lambda(x-y)}m_{12}\\
e^{2i\lambda(x-y)}m_{21} & m_{22}\\
\end{array}\right).
\end{align*}
Note that $e^{2i\lambda(x-y)}=e^{-\frac{|z|^{2}-q_{0}^{2}}{|z|^{2}}Imz(x-y)}e^{2iRez(x-y)}$, obviously the function $u_{-,1}$ is analytic in the region $(|z|^{2}-q_{0}^{2})Imz>0$, i.e., the region $D^{+}$.

\section*{Appendix B: Trace formula and theta condition}
The so-called trace formula uses the scattering data (discrete eigenvalues and reflection coefficient) to express the analytical scattering coefficients $s_{11}(z)$ and $s_{22}(z)$. In addition, under the condition of nonzero boundary, the trace formula can also provide the asymptotic phase difference of potential and scattering data. Now we first consider the discrete eigenvalues are simple zeros, namely, $s_{11}(z_{n})=0$ and $s'_{11}(z_{n})\neq0$. In terms of the previous analysis, we know that $s_{11}(z)$ is analytic in the region $D^{-}$ at $z=z_{n}$ and $z=-q_{0}^{2}/z_{n}^{*}$, and $s_{22}(z)$ is analytic in the region $D^{+}$ at $z=z_{n}^{*}$ and $z=-q_{0}^{2}/z_{n}$. We thus introduce the functions
\begin{align*}
\chi^{-}(z)=s_{11}(z)\prod_{n=1}^{2N}\frac{(z-z_{n}^{*})(z+q_{0}^{2}/z_{n})}
{(z-z_{n})(z+q_{0}^{2}/z_{n}^{*})},\\
\chi^{+}(z)=s_{22}(z)\prod_{n=1}^{2N}\frac{(z-z_{n})(z+q_{0}^{2}/z_{n}^{*})}
{(z-z_{n}^{*})(z+q_{0}^{2}/z_{n})},
\end{align*}
which are analytic in $D^{-}$ and $D^{+}$ respectively. In additional, they are not have zeros in the relevant analytical region. Resorting to Eq. \eqref{Sz}, as $z\rightarrow\infty$ we have $\chi^{\pm}\rightarrow 1$, further $\det S(z)=1$. Obviously
\begin{align*}
\chi^{+}(z)\chi^{-}(z)=\frac{1}{1+\rho(z)\rho^{*}(z^{*})},\quad z\in\Sigma,
\end{align*}
which can be converted to a scalar, multiplicative RH problem
\begin{align*}
\log\chi^{+}(z)+\log\chi^{-}(z)=
-\log[1+\rho(z)\rho^{*})(z^{*})], \quad z\in\Sigma.
\end{align*}
The trace formula can be written as via solving the scalar RH problem
\begin{align*}
s_{11}(z)&=exp\left(\frac{1}{2\pi i}\int_{\Sigma}\frac{\log[1+\rho(\zeta)\rho^{*}
(\zeta^{*})]}{\zeta-z}\,d\zeta\right)\prod_{n=1}^{2N}\frac{(z-z_{n})
(z+q_{0}^{2}/z_{n}^{*})}{(z-z_{n}^{*})(z+q_{0}^{2}/z_{n})},\\
s_{22}(z)&=exp\left(-\frac{1}{2\pi i}\int_{\Sigma}\frac{\log[1+\rho(\zeta)\rho^{*}
(\zeta^{*})]}{\zeta-z}\,d\zeta\right)\prod_{n=1}^{2N}\frac{(z-z_{n}^{*})
(z+q_{0}^{2}/z_{n})}{(z-z_{n})(z+q_{0}^{2}/z_{n}^{*})}.
\end{align*}
Note that as $z\rightarrow0$, $s_{11}(z)\rightarrow q_{-}/q_{+}$ from the Eq. \eqref{Sz}, one has
\begin{align*}
\prod_{n=1}^{2N}\frac{(z-z_{n}^{*})(z+q_{0}^{2}/z_{n})}
{(z-z_{n})(z+q_{0}^{2}/z_{n}^{*})}\rightarrow1,\quad z\rightarrow 0.
\end{align*}
The trace formula of $s_{11}(z)$ implies
\begin{align*}
q_{-}/q_{+}=exp\left(\frac{1}{2\pi i}\int_{\Sigma}
\frac{\log[1+\rho(\zeta)\rho^{*}(\zeta^{*})]}{\zeta}\,d\zeta\right),\quad for \quad z\rightarrow 0,
\end{align*}
therefore the theta conditions can be given
\begin{align*}
\arg\frac{q_{-}}{q_{+}}=\frac{1}{2\pi}\int_{\Sigma}
\frac{\log[1+\rho(\zeta)\rho^{*}(\zeta^{*})]}{\zeta}\,d\zeta+4\sum_{n=1}^{N}\arg z_{n}.
\end{align*}

For double zeros, namely $s_{11}(z_{n})=s'_{11}(z_{n})=0$ and $s''_{11}(z_{n})\neq0$, we can similarly be obtain the trace formulae and theta condition. Letting
\begin{align*}
\hat{\chi}^{-}(z)=s_{11}(z)\prod_{n=1}^{2N}\frac{(z-z_{n}^{*})^{2}(z+q_{0}^{2}/z_{n})^{2}}
{(z-z_{n})^{2}(z+q_{0}^{2}/z_{n}^{*})^{2}},\\
\hat{\chi}^{+}(z)=s_{22}(z)\prod_{n=1}^{2N}\frac{(z-z_{n})^{2}(z+q_{0}^{2}/z_{n}^{*})^{2}}
{(z-z_{n}^{*})^{2}(z+q_{0}^{2}/z_{n})^{2}}.
\end{align*}
Then similar to the simple zeros, the following trace formulas and theta condition can be obtained
\begin{align*}
s_{11}(z)&=exp\left(-\frac{1}{2\pi i }\int_{\Sigma}\frac{\log[1+\rho(\zeta)\rho^{*}(\zeta^{*})]}{\zeta-z}
\,d\zeta\right)\prod_{n=1}^{2N}\frac{(z-z_{n})^{2}(z+q_{0}^{2}/z_{n}^{*})^{2}}
{(z-z_{n}^{*})^{2}(z+q_{0}^{2}/z_{n})^{2}},\\
s_{22}(z)&=exp\left(\frac{1}{2\pi i }\int_{\Sigma}\frac{\log[1+\rho(\zeta)\rho^{*}(\zeta^{*})]}{\zeta-z}
\,d\zeta\right)\prod_{n=1}^{2N}\frac{(z-z_{n}^{*})^{2}(z+q_{0}^{2}/z_{n})^{2}}
{(z-z_{n})^{2}(z+q_{0}^{2}/z_{n}^{*})^{2}}.
\end{align*}
\begin{align*}
argq_{-}-argq_{+}=\frac{1}{2\pi}\int_{\Sigma}
\frac{\log[1+\rho(\zeta)\rho^{*}(\zeta^{*})]}{\zeta}\,d\zeta+8\sum_{n=1}^{N}\arg z_{n}.
\end{align*}

\section*{Acknowledgements}

This work was supported by the Postgraduate Research and Practice of Educational Reform for Graduate students in CUMT under Grant No. 2019YJSJG046, the Natural Science Foundation of Jiangsu Province under Grant No. BK20181351, the Six Talent Peaks Project in Jiangsu Province under Grant No. JY-059, the Qinglan Project of Jiangsu Province of China, the National Natural Science Foundation of China under Grant No. 11975306, the Fundamental Research Fund for the Central Universities under the Grant Nos. 2019ZDPY07 and 2019QNA35, and the General Financial Grant from the China Postdoctoral Science Foundation under Grant Nos. 2015M570498 and 2017T100413.


\begin{thebibliography}{3}
\bibitem{BBJ-2007}
J. Belmonte-Beitia,  V. M. P\'{e}rez-Garc\'{\i}a, V.  Vekslerchik,
 P. J. Torres, Lie symmetries and solitons in nonlinear systems with spatially inhomogeneous nonlinearities, Phys. Rev. Lett. 98 (2007) 064102.
\bibitem{PG-1998}
 V. M. Perez-Garcia,  H. Michinel,  H. Herrero, Bose-
Einstein solitons in highly asymmetric traps, Phys. Rev. A 57 (1998) 3837-3842.
\bibitem{RR-2007}
 R. Radha,  V. R. Kumar, Explode-Decay Solitons in the Generalized Inhomogeneous Higher-Order Nonlinear Schr\"{o}dinger Equations, Z. Naturforsch. A 62(7-8) (2007) 381-386.
\bibitem{FengLL-2019}
L. L. Feng,  S. F. Tian,  T. T. Zhang, Solitary wave, breather wave and rogue wave solutions of an inhomogeneous fifth-order nonlinear Schr\"{o}dinger equation from Heisenberg ferromagnetism, Rocky Mountain J. Math. 49(1) (2019) 29-45.
\bibitem{Yinnan-2015}
 C. Yinnan, Rogue wave solutions for an inhomogeneous fifth-order nonlinear Schr\"{o}dinger equation from Heisenberg ferromagnetism, J. Progress. Res. Math, 4(2) (2015) 328-338.
\bibitem{Shabat-1976}
A. B. Shabat,  One dimensional perturbations of a differential operator and the
inverse scattering problem, Problems in mechanics and mathematical physics, (1976)
279296.
\bibitem{Zakharov-1984}
V.E. Zakharov, S.V. Manakov, S.P. Novikov, L.P. Pitaevskii, The Theory of Solitons: The Inverse Scattering Method, Consultants Bureau, New York, 1984.


\bibitem{Ablowitz-1991}
M. J. Ablowitz, P. A. Clarkson, Solitons; Nonlinear Evolution Equations and In-
verse Scattering, Cambrige Univ. Press, 1991.
\bibitem{RHP-1}
W. X. Ma, Riemann-Hilbert problems and $N$-soliton solutions for a coupled mKdV system, J. Geom. Phys. 132 (2018) 45-54.
\bibitem{RHP-2}
S. F.  Tian, Initial-boundary value problems for the general coupled nonlinear Schr\"{o}dinger equation on the interval via the Fokas method, J. Differ. Equ.  262(1) (2017) 506-558.
\bibitem{RHP-3}
X. Geng, J. Wu, Riemann-Hilbert approach and $N$-soliton solutions for a generalized Sasa-Satsuma equation, Wave Motion.  60 (2016)  62-72.
\bibitem{RHP-4}
S. F. Tian, The mixed coupled nonlinear Schr\"{o}dinger equation on the half-line via the Fokas method, Proc. R. Soc. Lond. A 472(2195)  (2016)  20160588.
\bibitem{RHP-5}
D. S. Wang, D. J. Zhang,  J. Yang, Integrable properties of the general coupled nonlinear Schr\"{o}dinger equations, J. Math. Phys.  51(2) (2010) 023510.
\bibitem{RHP-7}
W. Q. Peng, S. F. Tian, X. B. Wang, T. T. Zhang, Y. Fang, Riemann-Hilbert method and multi-soliton solutions for three-component coupled nonlinear Schr\"{o}dinger equations, J. Geom. Phys. 146 (2019) 103508.
\bibitem{RHP-8}
 Y. Zhang,  Y. Cheng,  J. He, Riemann-Hilbert method and $N$-soliton for two-component Gerdjikov-Ivanov equation, J. Nonlinear Math. Phys.  24(2) (2017) 210-223.
\bibitem{RHP-9}
B. Guo,  N. Liu,  Y. Wang, A Riemann-Hilbert approach for a new type coupled nonlinear Schr\"{o}dinger equations, J. Math. Anal. Appl.  459(1) (2018) 145-158.

\bibitem{RHP-11}
J. J. Yang, S. F. Tian, W. Q. Peng, T. T. Zhang, The $N$-coupled higher-order nonlinear Schr\"{o}dinger equation: Riemann-Hilbert problem and multi-soliton solutions, Math. Meth. Appl. Sci. https://doi.org/10.1002/mma.6055.
\bibitem{longtime-0}
P. Deift,  X. Zhou, A steepest descent method for oscillatory Riemann-Hilbert problems. Asymptotics for the MKdV equation, Ann. Math. 137(2) (1993) 295-368.
\bibitem{longtime-1}
J. Xu, E. Fan, Long-time asymptotics for the Fokas-Lenells equation with decaying initial value problem: without solitons, J. Differ. Equ.  259(3) (2015) 1098-1148.

\bibitem{longtime-4}
D. S. Wang, X.  Wang, Long-time asymptotics and the bright $N$-soliton solutions of the Kundu-Eckhaus equation via the Riemann-Hilbert approach, Nonlinear Anal. Real World Appl.  41 (2018) 334-361.
\bibitem{TZ-2018}
S.F. Tian, T.T. Zhang, Long-time asymptotic behavior for the Gerdjikov-Ivanov type of derivative nonlinear Schr\"{o}dinger equation with time-periodic boundary condition, Proc. Amer. Math. Soc. 146 (4) (2018) 1713-1729.
\bibitem{longtime-5}
B. Guo,  N. Liu, Long-time asymptotics for the Kundu-Eckhaus equation on the half-line, J. Math. Phys.  59(6) (2018) 061505.
\bibitem{longtime-7}
H. Liu,  X. Geng,  B. Xue, The Deift-Zhou steepest descent method to long-time asymptotics for the Sasa-Satsuma equation, J. Differ. Equ.  265(11) (2018) 5984-6008.
\bibitem{NZBC-1}
M. J. Ablowitz, X. D. Luo, Z. H. Musslimani, Inverse scattering transform for the nonlocal nonlinear Schr\"{o}dinger equation with nonzero boundary conditions, J. Math. Phys. 59(1) (2018) 011501.
\bibitem{NZBC-2}
M. J. Ablowitz, B. F. Feng, X. D. Luo,  Z. H. Musslimani, Reverse Space-Time Nonlocal Sine-Gordon/Sinh-Gordon Equations with Nonzero Boundary Conditions, Stud. Appl. Math. 141(3) (2018) 267-307.
I\bibitem{Yangjj2} J.J. Yang, S.F. Tian, Z.Q. Li, Inverse scattering transform and soliton solutions for the focusing Kundu-Eckhaus equation with nonvanishing boundary conditions,  arXiv:1911.00340.
\bibitem{NZBC-3}
G. Zhang, S. Chen, Z. Yan, Focusing and defocusing Hirota equations with non-zero boundary conditions: Inverse scattering transforms and soliton solutions, Commun. Nonlinear Sci. Numer. Simul. 80 (2020)  104927.
\bibitem{NZBC-4}
J. Zhu,  L. Wang,  X. Geng, Riemann-Hilbert approach to TD equation with nonzero boundary condition, Front. Math. China  13(5) (2018) 1245-1265.
\bibitem{NZBC-5}
J.J. Yang, S.F. Tian, Riemann-Hilbert problem for the modified Landau-Lifshitz equation with nonzero boundary conditions, arXiv:1909.11263.
\bibitem{NZBC-li}
Z.Q. Li, S.F. Tian, J.J. Yang, Riemann-Hilbert approach and soliton solutions for the higher-order dispersive nonlinear Schr\"{o}dinger equation with nonzero boundary conditions, arXiv:1911.01624.
\bibitem{NZBC-mao}
 J. J. Mao, S. F.Tian, Rieman-Hilbert approach for the NLSLab equation with nonzero boundary conditions, arXiv:1911.00683.
\bibitem{NZBC-6}
B. Prinari, M. J. Ablowitz, and G. Biondini, Inverse scattering transform for the vector nonlinear Schr\"{o}dinger equation with nonvanishing boundary conditions, J. Math. Phys. 47 (2006) 063508.
\bibitem{NZBC-7}
M. J. Ablowitz, G. Biondini, and B. Prinari, Inverse scattering transform for the integrable discrete nonlinear Schr\"{o}dinger equation with nonvanishing boundary conditions, Inverse Prob. 23 (2007) 1711-1758.
\bibitem{NZBC-8}
B. Prinari, G. Biondini, and A. D. Trubatch, Inverse scattering transform for the multi-component nonlinear Schr\"{o}dinger equation with nonzero boundary conditions, Stud. Appl. Math. 126 (2011) 245-302.
\bibitem{NZBC-9}
F. Demontis, B. Prinari, C. van der Mee, and F. Vitale, The inverse scattering transform for the defocusing
nonlinear Schr\"{o}dinger equations with nonzero boundary conditions, Stud. Appl. Math. 131 (2013) 1-40.
\bibitem{NZBC-10}
B. Prinari and F. Vitale, Inverse scattering transform for the focusing nonlinear Schr\"{o}dinger equation with one-sided nonzero boundary condition, Cont. Math. 651 (2015) 157-194.
\bibitem{NZBC-13}
G. Biondini,  D. Kraus, Inverse scattering transform for the defocusing Manakov system with nonzero boundary conditions, SIAM J. Math. Anal.  47(1) (2015) 706-757.
\bibitem{NZBC-14}
G. Biondini,  G. Kova\u{c}i\u{c}, Inverse scattering transform for the focusing nonlinear Schr\"{o}dinger equation with nonzero boundary conditions, J. Math. Phys.  55(3) (2014) 031506.
\bibitem{NZBC-15}
Y. Yang, E. Fan, Riemann-Hilbert approach to the modified nonlinear Schr\"{o}dinger equation with non-vanishing asymptotic boundary conditions, arXiv:1910.07720.
\bibitem{NZBC-16}
L. Wen, E. Fan, The Sasa-Satsuma equation with non-vanishing boundary conditions, arXiv:1911.11944.
\bibitem{NZBCtime-1}
G. Biondini, D. Mantzavinos, Long-time asymptotics for the focusing nonlinear Schr\"{o}dinger equation with nonzero boundary conditions at infinity and asymptotic stage of modulational instability, Comm. Pure Appl. Math. 70(12) (2017) 2300-2365.
\bibitem{NZBCtime-2}
G. Biondini, S. Li,  D. Mantzavinos, Long-time asymptotics for the focusing nonlinear Schr\"{o}dinger equation with nonzero boundary conditions in the presence of a discrete spectru, arXiv:1907.09432.
\bibitem{NZBCtime-3}
D. S. Wang, B. Guo, X. Wang, Long-time asymptotics of the focusing Kundu-Eckhaus equation with nonzero boundary conditions, J. Different. Equ. 266(9) (2019) 5209-5253.
\bibitem{NZBCtime-4}
W. Q. Peng,  S. F. Tian, Long-time asymptotics in the modified Landau-Lifshitz equation with nonzero boundary conditions, arXiv:1912.00542.
\bibitem{NZBCtime-5}
B. Guo, N. Liu, The Gerdjikov-Ivanov-type derivative nonlinear Schr\"{o}dinger equation: Long-time dynamics of nonzero boundary conditions, Math. Meth. Appl. Sci. 42(14)  (2018) 4839-4861.
\bibitem{shock-2007}
R. Buckingham, S. Venakides, Long-time asymptotics of the nonlinear Schr\"{o}dinger equation Shock problem, Comm. Pure Appl. Math. LX 60(9) (2007) 1349-1414.
\bibitem{step-2011}
A. B. De Monvel, V. P. Kotlyarov, D. Shepelsky, Focusing NLS equation: Long-time dynamics of step-like initial data, Int. Math. Res. Not. IMRN 2011(7) (2011) 1613-1653.
\bibitem{Faddeev-1987}
L. D. Faddeev, L. A. Takhtajan, Hamiltonian Methods in the Theory of Solitons Springer, Berlin, 1987.
\bibitem{Liu}
Liouville's formula, wikipedia,
https://en.wikipedia.org/wiki/Liouville$\%$27$s\_$formula.

\end{thebibliography}
\end{document}